\PassOptionsToPackage{unicode}{hyperref}
\PassOptionsToPackage{hyphens}{url}
\PassOptionsToPackage{dvipsnames,svgnames,x11names,table}{xcolor}

\documentclass[9pt,arxiv]{templates/lapreprint}
\usepackage[version=4]{mhchem} % For chemical notation
\usepackage{siunitx}    % For SI units
\usepackage{pdflscape}  % For putting pages in landscape mode
\usepackage{rotating}   % For rotating specific elements
\usepackage{textgreek}  % Greek symbols
\usepackage{gensymb}    % Symbols
\usepackage[misc]{ifsym} % For the \Letter symbol
\usepackage{listings}   % For inserting code chunks
\usepackage{colortbl}   % For Knitr table colouring
\usepackage{tabularx}   % For making Knitr tables compatible
\usepackage{subcaption}
\usepackage{multirow}
\usepackage{snotez}     % For sidenote environments. enotez for endnotes
\DeclareSIUnit\Molar{M}

\usepackage{hyperref} % provides \hypertarget
\usepackage{graphicx}

\makeatletter
\def\maxwidth{\ifdim\Gin@nat@width>\linewidth\linewidth\else\Gin@nat@width\fi}
\def\maxheight{\ifdim\Gin@nat@height>\textheight\textheight\else\Gin@nat@height\fi}
\makeatother
\setkeys{Gin}{width=\maxwidth,height=\maxheight,keepaspectratio}
\makeatletter
\def\fps@figure{htbp}
\makeatother

\usepackage{etoolbox}
\BeforeBeginEnvironment{longtable}{\begin{center}\scriptsize}
\AfterEndEnvironment{longtable}{\end{center}}

\newlength{\cslhangindent}
\newlength{\csllabelwidth}
\newenvironment{CSLReferences}[2] % #1 hanging-ident, #2 entry spacing
 {% don't indent paragraphs
  \setlength{\parindent}{0pt}
  \ifodd #1 \everypar{\setlength{\hangindent}{\cslhangindent}}\ignorespaces\fi
  \ifnum #2 > 0
  \setlength{\parskip}{#2\baselineskip}
  \fi
 }%
 {}
\usepackage{calc}

\usepackage{longtable}
\usepackage{booktabs}
\usepackage{tabularx}
\usepackage{calc}

\title{Feel Old Yet? Updating Mode of Transportation Distributions from
Travel Surveys using Data Fusion with Mobile Phone Data}

\author[1,2]{Eduardo Graells-Garrido}

\author[3]{Daniela Opitz}

\author[4]{Francisco Rowe}

\author[5]{Jacqueline Arriagada}

\affil[1]{Department of Computer Science, University of Chile}
\affil[2]{National Center for Artificial Intelligence (CENIA)}
\affil[3]{Data Science Institute, Faculty of Engineering, Universidad
del Desarrollo}
\affil[4]{Department of Geography and Planning, University of Liverpool}
\affil[5]{University of Leeds}

\metadata[]{\Letter\hspace{.5ex} For correspondence}{\url{{egraells@dcc.uchile.cl}}}

\metadata[]{Data availability}{The Telefónica Movistar mobile phone
records have been obtained directly from the mobile phone operator
through an agreement between the Data Science Institute from Universidad
del Desarrollo and Telefónica R\&D. This mobile phone operator retains
ownership of these data and imposes standard provisions to their sharing
and access which guarantee privacy. Anonymized datasets are available
from Telefónica R\&D Chile for researchers who meet the criteria for
access to confidential data. The other datasets used in this study are
publicly available.}

\metadata[]{Acknowledgments}{We thank GPA Movistar for facilitating the
data for this study, and Leo Ferres for insightful discussion. E.
Graells-Garrido was partially funded by ANID Fondecyt \#1211490 and National Center for Artificial Intelligence CENIA FB210017, Basal ANID. D.
Opitz was partially funded by ANID \#PAI77190057 and ANID Fondecyt de
Iniciación \#11220799.}

\metadata[]{Competing interests}{The author declare no competing interests.}

\leadauthor{Graells-Garrido}
\shorttitle{Updating Mode of Transportation Distributions from Travel
Surveys using Data Fusion}

\usepackage{libertinus}

\begin{document}
\maketitle

\begin{abstract}
Up-to-date information on different modes of travel to monitor transport
traffic and evaluate rapid urban transport planning interventions is
often lacking. Transport systems typically rely on traditional data
sources providing outdated mode-of-travel data due to their data
latency, infrequent data collection and high cost. To address this
issue, we propose a method that leverages mobile phone data as a
cost-effective and rich source of geospatial information to capture
current human mobility patterns at unprecedented spatiotemporal
resolution. Our approach employs mobile phone application usage traces
to infer modes of transportation that are challenging to identify (bikes
and ride-hailing/taxi services) based on mobile phone location data.
Using data fusion and matrix factorization techniques, we integrate
official data sources (household surveys and census data) with mobile
phone application usage data. This integration enables us to reconstruct
the official data and create an updated dataset that incorporates
insights from digital footprint data from application usage. We
illustrate our method using a case study focused on Santiago, Chile
successfully inferring four modes of transportation: mass-transit (all
public transportation), motorised (cars and similar vehicles), active
(pedestrian and cycle trips), and taxi (traditional taxi and
ride-hailing services). Our analysis revealed significant changes in
transportation patterns between 2012 and 2020. We quantify a reduction
in mass-transit usage across municipalities in Santiago, except where
metro/rail lines have been more recently introduced, highlighting added
resilience to the public transport network of these infrastructure
enhancements. Additionally, we evidence an overall increase in motorised
transport throughout Santiago, revealing persistent challenges in
promoting urban sustainable transportation. Findings also point to a
rise in the share of taxi usage, and a drop in active mobility,
suggesting a modal shift towards less sustainable modes of travel. We
validate our findings comparing our updated estimates with official
smart card transaction data. The consistency of findings with expert
domain knowledge from the literature and historical transport usage
trends further support the robustness of our approach.
\end{abstract}

\hypertarget{introduction}{%
\section{Introduction}\label{introduction}}

Understanding travel behaviour is key to creating more resilient and
sustainable urban transport systems. The global collective contribution
of the transport sector to carbon emissions was estimated to be over
14\% of the global estimate
(\protect\hyperlink{ref-policymakersbook2012}{Allen et al., 2012}). As
cities expand and absorb population growth, global warming
(\protect\hyperlink{ref-acar2020potential}{Acar and Dincer, 2020}), air
pollution (\protect\hyperlink{ref-colvile2001transport}{Colvile et al.,
2001}), traffic congestion
(\protect\hyperlink{ref-thomson1998reflections}{Thomson, 1998}), and
urban sprawl (\protect\hyperlink{ref-brueckner2005transport}{Brueckner,
2005}) exert increasing pressure on decarbonising transport networks and
prioritising more sustainable modes of mobility.

However, cities often lack appropriate data analytics to evaluate and
implement transport planning interventions to achieve these goals.
Traditional data sources, such as travel surveys, manual count studies,
censuses and stated preference surveys have conventionally been used to
estimate traffic counts, potential demand for transport and assess the
spatial patterns of human mobility within cities
(\protect\hyperlink{ref-zannat2019emerging}{Zannat and Choudhury,
2019}). Yet, these data sources are expensive, infrequent and suffer
from data latency (\protect\hyperlink{ref-rowe2023big}{Rowe, 2023}).
Thus, while these data from traditional sources may provide an accurate
representation of urban transport networks at a given time, such
representations may be unreliable or outdated a few years into the
future, particularly in fast-expanding cities or those that experience
shocks due to social unrest, natural hazards
(\protect\hyperlink{ref-rowe2022using}{Rowe, 2022}) or pandemics
(\protect\hyperlink{ref-green2021new}{Green et al., 2021}).

Mobility information obtained from sources such as the Global System for
Mobile Communication (GSM) network can help overcome these deficiencies
(\protect\hyperlink{ref-rowe2023big}{Rowe, 2023}). Mobile network data,
when they are available, provide a cost-effective source of geospatial
information to capture human mobility at unprecedented geographic and
temporal granularity (\protect\hyperlink{ref-zannat2019emerging}{Zannat
and Choudhury, 2019}). Mobile phone data also offer information to
capture travel behaviour in real or near-real time, providing an
opportunity to frequently update information about how people move
around cities to support transport planning
(\protect\hyperlink{ref-antoniou2011synthesis}{Antoniou et al., 2011}).
However, they only provide geospatial information . They do not deliver
data on users' personal attributes, such as age, gender, and income
which are often associated with different travel behaviours, or
information on mode of transportation. Information on modal split is
however a key ingredient to developing transportation interventions to
reduce private car usage and promote sustainable transport modes
(\protect\hyperlink{ref-rodriguez2014measuring}{Rodríguez-Núñez and
García-Palomares, 2014}).

To leverage the strengths and address these limitations of traditional
and new emerging forms of data, we develop a generalisable Data Fusion
(DF) framework to update area-level estimates of mode of transport trip
counts (defined as mode split). DF involves the integration of multiple
related datasets to render a unified representation that facilitates the
identification of relevant patterns.

In transportation, DF has been used for multiple purposes, such as
traffic demand prediction and forecasting
(\protect\hyperlink{ref-el2011data}{El Faouzi et al., 2011}). Unlike
prior work, we use DF to produced updated mode split estimates from
travel survey data by integrating inferred origin-destination mode of
transport distributions from more up-to-date mobile phone data from Deep
Packet Inspection (DPI), Extended Detail Records (XDR), and
socioeconomic and demographic information from a representative
household survey and census data. XDR and DPI data are used to infer
more up-to-date origin-destination mode of transport estimates. XDRs
provide geospatial locations of users, and DPI data offer information on
application usage of related transport modes to infer these estimates
based on existing travel survey data. Household survey and census data
are used to incorporate data on the socio-economic and demographic
composition of the resident population and adjust biases for
distributional biases relating to these attributes in mobile phone data.

Our work contributes to expanding the approach followed by
Graells-Garrido et al.
(\protect\hyperlink{ref-graells2018inferring}{2018b}), to estimate the
modal split but using a DF framework by incorporating data: (1) from
mobile phone application records to infer the use of bike-sharing
platforms and taxi services to improve estimates of bicycle and taxi
transport modes; and (2) from socioeconomic and demographic attributes
of the local resident population to integrate information about
variation in the usage of different transport modes.

This paper is structured as follows. The Background section delves
deeper into the existing literature and research related to transport
data collection and analysis methods regarding the modal split. In the
Context and Data section, we present the specific context of our study,
focusing on Santiago, Chile, and describe the data sources used,
including household surveys, census data, and mobile phone application
usage traces. The Methods section outlines our approach, which involves
leveraging mobile phone location data and employing data fusion and
matrix factorization techniques to integrate official data sources with
digital footprint data. In the Results section, we present the findings
of our analysis, including insights into transportation patterns and
shifts between 2012 and 2020 in Santiago, and a thorough analysis and
interpretation of the results, comparing them with expert domain
knowledge and validation data. The Discussion section explains the
practical and theoretical implications of our work, as well as its
limitations and future lines of research. Finally, the Conclusions
section closes the paper with a highlight of our contribution to
knowledge and transportation planning.

\hypertarget{background}{%
\section{Background}\label{background}}

Information on transportation modes is critical to urban transport
planning. It enables improving the efficiency of transport systems by
identifying suitable options for specific travel needs (e.g.,
long-distance commuting or freight transport) to design multimodal
transportation networks (e.g., Karlaftis
(\protect\hyperlink{ref-karlaftis2004dea}{2004})). Accurate information
on modes of transport also enables assessment of transport demand by
identifying locations of high demand for specific travel options to
inform the appropriate allocation of resources
(\protect\hyperlink{ref-karlaftis2004dea}{Karlaftis, 2004}). Different
modes of transport require different types of infrastructure and have
varying environmental footprint (e.g., Vuchic
(\protect\hyperlink{ref-vuchic2007urban}{2007})). Having current
information on modes of transport is thus essential to identify critical
infrastructure needs and develop strategies to reduce carbon emissions
and promote more sustainable forms of urban mobility
(\protect\hyperlink{ref-vuchic2007urban}{Vuchic, 2007}).

The estimation of transportation modes using traditional statistical
methods and machine learning techniques is a well-established problem in
transportation research. Traditionally, estimates of mode split have
been derived from data collected through household travel or travel
diary panel surveys, involving retrospective, self-reporting records of
how often people use various modes of transport
(\protect\hyperlink{ref-lee2020millennials}{Lee et al., 2020};
\protect\hyperlink{ref-molin2016multimodal}{Molin et al., 2016}), or of
individuals' trips for a selected number of days
(\protect\hyperlink{ref-heinen2015same}{Heinen and Chatterjee, 2015};
\protect\hyperlink{ref-kuhnimhof2006users}{Kuhnimhof et al., 2006}).
Data collection through these surveys is however costly and
time-consuming posing significant limitations to their scalability to
cover large geographical areas, population samples and complex
multimodal systems. In recognition of these limitations, increasing
efforts have been made to leverage on the emergence of digital footprint
data to estimate transportation modes (e.g., Graells-Garrido et al.
(\protect\hyperlink{ref-graells2018inferring}{2018b}), Bachir et al.
(\protect\hyperlink{ref-bachir2019mode}{2019})).

Two forms of digital footprint data feature prominently in the
estimation of travel counts by transportation modes: (1) mobile network
data (including DPI and XDRs), and (2) GPS data. GPS data have been used
to estimate mode of transport use by leveraging on the accuracy of GPS
technology. They provide accurate location information to render
individual travel trajectories in real time by triangulating records on
speed, time and location (\protect\hyperlink{ref-bantis2017you}{Bantis
and Haworth, 2017}; \protect\hyperlink{ref-dabiri2018inferring}{Dabiri
and Heaslip, 2018}). However, the availability of individual GPS records
is ethically sensitive, geographically sparse and limited to a small set
of users as their collection relies on the use of tracking mobile phone
applications (\protect\hyperlink{ref-roweetal2022}{Rowe et al., 2022}).
Only a segment of the population uses a given application or allows
tracking (\protect\hyperlink{ref-grantz2020use}{Grantz et al., 2020}).
Tech companies (e.g., Apple) have included features to prevent tracking
by mobile phone applications and new legislation (e.g., GDPR) has
tightened the use of mobile phones to collect and store personal data.
All these issues challenge the use of GPS records to accurately capture
human mobility and hence derive reliable mode split estimates.

In contrast, mobile network data generally cover entire geographical
areas (e.g.~countries) and population systems (i.e.~users of a network
operator). They are also a by-product of an administrative process and
thus continuously generated by network operators. As a result, mobile
network data offer more extensive coverage and are more readily
available than GPS data. However, mobile network data are geographically
sparse and irregular as their collection depends on the location and
facing position of telecommunication antennas. They do not provide exact
information on location. Users' location needs to be inferred from their
position relative to telecommunication antennas. This problem represents
a challenge to accurately estimate the use of different modes of
transportation at a given location.

A large number of studies have focused on the use of mobile network data
to address this challenge and estimate the use of different modes of
transportation(\protect\hyperlink{ref-huang2019transport}{Huang et al.,
2019}). A large body of the literature has focused on estimating the use
of trains and cars, particularly using data on intercity trips
(\protect\hyperlink{ref-doyle2011utilising}{Doyle et al., 2011};
\protect\hyperlink{ref-garcia2016big}{García et al., 2016};
\protect\hyperlink{ref-hui2017investigating}{Hui et al., 2017};
\protect\hyperlink{ref-schlaich2010generating}{Schlaich et al., 2010};
\protect\hyperlink{ref-smoreda2013spatiotemporal}{Smoreda et al., 2013};
\protect\hyperlink{ref-wu2013studying}{Wu et al., 2013}). The focus on
intercity trips may have been determined by the accuracy of the
estimates that can be produced inferring train and car trips from
relatively low spatial and temporal resolution mobile phone network
data. Less work has focused on inferring trips by alternative modes of
transportation, such as bus, trams, bikes and walk
(\protect\hyperlink{ref-graells2018inferring}{Graells-Garrido et al.,
2018b}; \protect\hyperlink{ref-holleczek2015traffic}{Holleczek et al.,
2015}; \protect\hyperlink{ref-horn2015deriving}{Horn and Kern, 2015};
\protect\hyperlink{ref-li2017estimating}{Li et al., 2017};
\protect\hyperlink{ref-wu2013studying}{Wu et al., 2013}). Alternative
modes of transportation are often aggregated into general groupings,
such as public transport versus private transport
(\protect\hyperlink{ref-horn2017qztool}{Horn et al., 2017};
\protect\hyperlink{ref-phithakkitnukoon2017inferring}{Phithakkitnukoon
et al., 2017}; \protect\hyperlink{ref-qu2015transportation}{Qu et al.,
2015}; \protect\hyperlink{ref-wang2010transportation}{Wang et al.,
2010}), air versus ground
(\protect\hyperlink{ref-hui2017investigating}{Hui et al., 2017}), moving
versus stationary (\protect\hyperlink{ref-calabrese2011}{Calabrese et
al., 2011}) or rail versus road
(\protect\hyperlink{ref-asgari2016inferring}{Asgari, 2016};
\protect\hyperlink{ref-bachir2019mode}{Bachir et al., 2019};
\protect\hyperlink{ref-doyle2011utilising}{Doyle et al., 2011}). While
useful in specific settings, such aggregations have more limited utility
in the current policy context seeking to develop planning interventions
to promote active forms of mobility, identify associated critical
infrastructure and thus reduce carbon emissions.

Previous work seeking to infer mode of transportation from mobile
network data has largely used non-supervised or semi-supervised machine
learning algorithms. Notable studies include Bachir et al.
(\protect\hyperlink{ref-bachir2019mode}{2019}), Graells-Garrido et al.
(\protect\hyperlink{ref-graells2018inferring}{2018b}), Kalatian and
Shafahi (\protect\hyperlink{ref-kalatian2016travel}{2016}), Wang et al.
(\protect\hyperlink{ref-wang2010transportation}{2010}). Wang et al.
(\protect\hyperlink{ref-wang2010transportation}{2010}) and Kalatian and
Shafahi (\protect\hyperlink{ref-kalatian2016travel}{2016}) employed
non-supervised machine learning algorithms and mobile phone data to
cluster travel times and travel speeds to identify different modes of
transport. Wang et al.
(\protect\hyperlink{ref-wang2010transportation}{2010}) identified two
forms of transport (i.e., road and public transport), while Kalatian and
Shafahi (\protect\hyperlink{ref-kalatian2016travel}{2016}) distinguished
between walking, public transportation and private car. Using data from
Santiago (Chile), Graells-Garrido et al.
(\protect\hyperlink{ref-graells2018inferring}{2018b}) proposed a method
based on semi-supervised Non-negative Matrix Factorization (NMF) to
various transport modes including: rail, car, bus, rail \& bus and
pedestrian modes. Bachir et al.
(\protect\hyperlink{ref-bachir2019mode}{2019}) employed agglomerative
hierarchical clustering to group Voronoi sectors of Paris with similar
transport usage and identify rail, road, and multimodal (rail+road)
modes. A novelty of their work was to use census data to rescale the
resulting origin-destination estimates to improve their statistical
representation.

While these studies have progressed our capacity to infer mode split
patterns from mobile network data, key challenges remain. For example,
Wang et al. (\protect\hyperlink{ref-wang2010transportation}{2010}) and
Kalatian and Shafahi (\protect\hyperlink{ref-kalatian2016travel}{2016})
only distinguished a selected number of transport modes and are based
solely on travel times or speed travel, which can vary at different
traffic stages and show similarities under certain conditions,
particularly in congested cities. The model proposed by Graells-Garrido
et al. (\protect\hyperlink{ref-graells2018inferring}{2018b}) can infer
multiple modes, including road, public transport, rail, pedestrian, and
rail+bus, but it tends to underestimate pedestrian mode results. On the
other hand, the model by Bachir et al.
(\protect\hyperlink{ref-bachir2019mode}{2019}) considers a bimodal
partition (rail and road) and does not account for the intermodality of
trips.

We argue that borrowing information from disparate data sources is key
to inferring different modes of transportation. NMF or traditional
clustering methods may not be the most suitable approaches to integrate
different data sources as they do not directly take advantage of
systematic relationships identified between attributes across multiple
datasets. We argue that data fusion algorithms can provide a more robust
approach to integrate disparate data sources as they are explicitly
designed to use systematic patterns observed across various datasets to
fuse them. This fusion is achieved by quantifying and identifying
mathematical patterns in the dataset as if they were part of a single
model. Data fusion algorithms have been predominantly applied in
biomedical research (\protect\hyperlink{ref-li2018review}{Li et al.,
2018}) and the modelling of urban lifestyles
(\protect\hyperlink{ref-xu2019mining}{González, 2019}). Their
application in urban transportation contexts is less prominent although
not absent (\protect\hyperlink{ref-el2011data}{El Faouzi et al., 2011}).
In our context, they have been used with travel surveys to update modal
split estimates (\protect\hyperlink{ref-chang2019modechoice}{Chang et
al., 2019}). However, they have not yet been employed to integrate
mobile phone data. We however recognise that the concept of data
integration is not new and has been incorporated in supervised
algorithms of modal split estimation by Huang et al.
(\protect\hyperlink{ref-huang2018modeling}{2018}). They used it to
combine mobile phone signalling data with subway smart card data and
taxi GPS data to develop a machine-learning model that predicts one-hour
ahead traffic population counts by area in Shenzhen (China). We also
recognise a variety of applications of supervised machine learning
approaches to estimate modal split based on mobile phone data. These
include Chen et al. (\protect\hyperlink{ref-chen2017mago}{2017}),
Semanjski et al. (\protect\hyperlink{ref-semanjski2017spatial}{2017})
and Breyer et al. (\protect\hyperlink{ref-breyer2022}{2022}). Chen et
al. (\protect\hyperlink{ref-chen2017mago}{2017}) uses the magnetic
signal combined with the accelerometer presented in intelligent devices
of different trips labelled with seven modes (stationary, bus, bike,
car, train, light rail and scooter). Semanjski et al.
(\protect\hyperlink{ref-semanjski2017spatial}{2017}) uses labelled trips
collected through the application Routecoach combined with urban
infrastructure obtained from OpenStreetMap (OSM), to predict the modes
of travel: driving, public transport, biking, and walking in the Flemish
Brabant province (Belgium). On the other hand, Breyer et al.
(\protect\hyperlink{ref-breyer2022}{2022}) proposes a semi-supervised
model to classify trips between cities from mobile network data. A key
limitation of supervised machine learning approaches is that they
require labelled data identifying the modes of travel. Such a labelled
dataset is difficult to generate at scale because they are costly and
normally cover a reduced geographical area.

Thus, significant progress has been made to generate area-level
estimates of transportation modes and to identify various modes of
transportation based on travel surveys and mobile phone data. Most
existing work has relied on non-supervised or semi-supervised machine
learning algorithms due to the limited availability of labelled data
identifying transportation modes. Less progress has, however, been made
in inferring trips undertaken by bike and emerging forms of shared
mobility, such as ride-hailing services. And, while data fusion
approaches have displayed great potential to integrate disparate
datasets and identify systematic patterns in order to infer modal split,
they have not been widely adopted to produce modes of transportation
estimates. This paper seeks to propose a data fusion framework to
integrate various data sources (list data sources) to estimate the
extent of bike and ride-hailing service usage, along with other public
and motorised forms of mobility.

\hypertarget{context-and-data}{%
\section{Context and Data}\label{context-and-data}}

Our work studies the Great Metropolitan Area of Santiago, Chile.
Santiago is a city with almost 8 million inhabitants and an urban
surface of 867.75 square kilometres as of 2017, composed of more than 40
independent municipalities, 35 within the urban area, whereas the rest
are on the periphery. The city has an integrated multimodal system with
an almost flat fare between metro, urban buses, and one rail service,
allowing up to three legs within a two-hour window. The \emph{bip!}
smart card is the only accepted payment method in the public
transportation system. The city also has ubiquitous taxi services and
access to several ride-hailing applications, and a public bike-sharing
system. For this work, we congregate all modes of transportation into
four categories: \emph{mass-transit}, which includes all public
transportation that uses the city smart card; \emph{motorised}, which
refers to cars and similar vehicles; \emph{active}, which includes
pedestrian and cycle trips; and \emph{taxi}, which includes traditional
taxi and ride-hailing applications.

The last official mode split for Santiago was published in 2015, based
on a travel survey from 2012. The survey is outdated considering the
different interventions and changes in the city, making it difficult to
understand current travel behaviour patterns. We emphasise three of
these changes. First, the population composition of the city has
significantly changed due to external and internal migration waves.
Second, new transportation infrastructure was built, including two metro
lines and a new train service from 2017 onwards, and new mobile phone
applications and other transportation systems appeared, including
ride-hailing apps such as Uber, electric scooters such as Lime, and
shared bike-systems such as Mobike. And third, the social upheaval from
October 2019 caused a series of massive demonstrations and multiple
riots throughout the city, including burned metro stations
(\protect\hyperlink{ref-urquieta2019metro}{Urquieta Ch., 2019}), the
closing of commercial districts, and an associated economic crisis
(\protect\hyperlink{ref-somma2021no}{Somma et al., 2021}).

To estimate an updated mode split, we integrated the following data
sources about Santiago: mobile phone data from 9---13 March 2020,
official data sources from 2020, urban infrastructure in 2020, and the
travel survey collected in 2012. We focus on trips made during morning
peak periods (between 6AM and 9AM) on weekdays. In this way, we try to
capture regular activity-based trips, such as work or study. On the
other hand, the morning peak period is the most congested, and therefore
the most interesting period from a planning perspective.

\hypertarget{urban-infrastructure-from-openstreetmap-osm}{%
\subsection{Urban Infrastructure from OpenStreetMap
(OSM)}\label{urban-infrastructure-from-openstreetmap-osm}}

\begin{figure}
\hypertarget{fig:osm}{%
\centering
\includegraphics{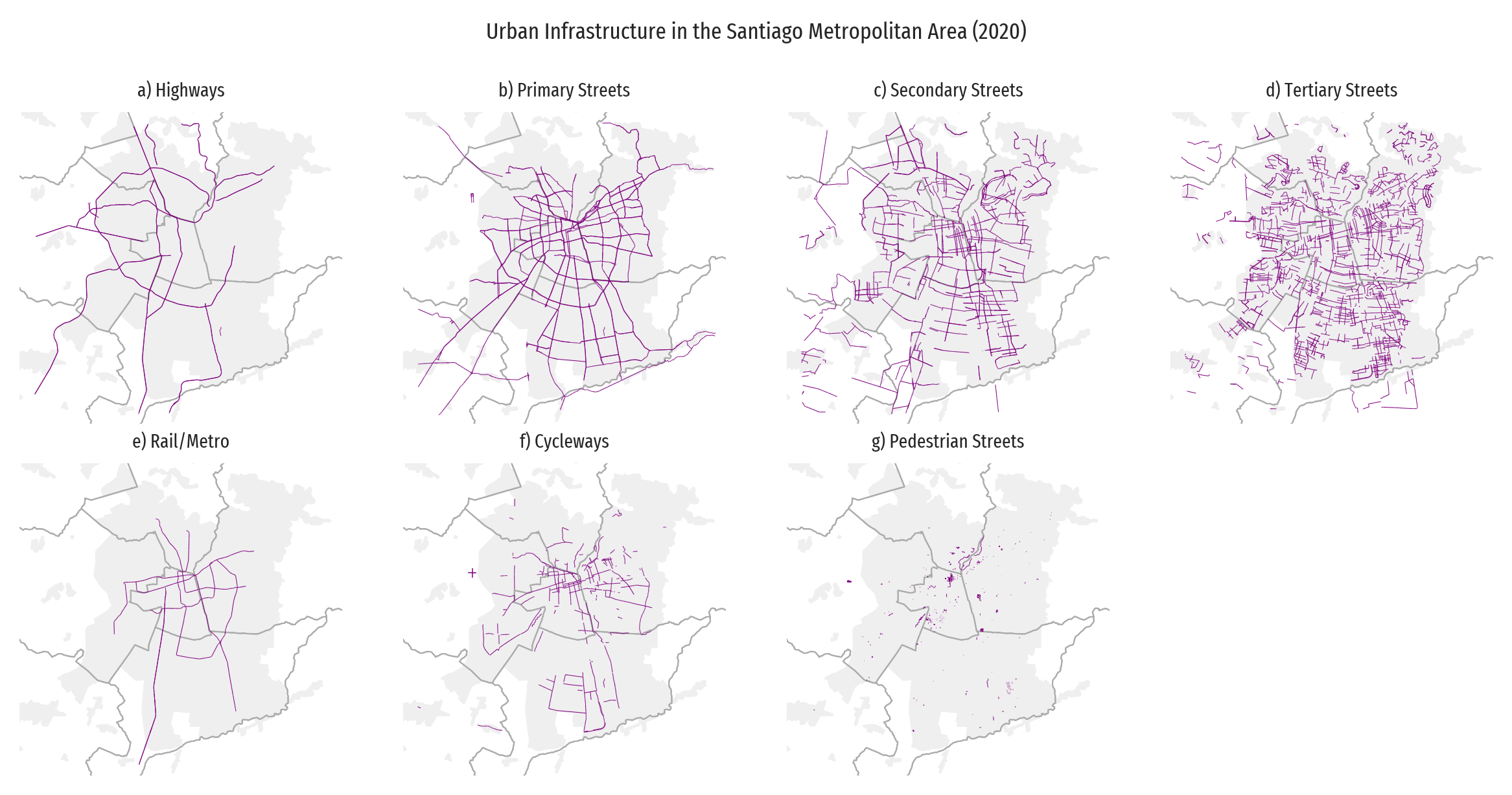}
\caption{Urban infrastructure from OpenStreetMap in March
2020.}\label{fig:osm}
}
\end{figure}

OSM is a global, open geographical database that is freely available and
fairly accurate for many cities
(\protect\hyperlink{ref-haklay2010good}{Haklay, 2010};
\protect\hyperlink{ref-zhang2019using}{Zhang and Pfoser, 2019}). It
contains several types of geographical features, including urban
infrastructure networks. OSM data can be downloaded freely, either from
the OSM organisation itself or in data aggregators such as GeoFabrik. We
parsed OSM data from March 2020 in Santiago using the pyrosm tool
(\protect\hyperlink{ref-pyrosm}{Tenkanen and contributors, 2022}). Then,
we identified the following urban network information: highways, primary
(or main) streets, secondary streets, tertiary streets, rail and metro
networks, cycleways, and pedestrian streets (see Figure \ref{fig:osm}).
These features are arguably associated with several modes of
transportation. We assume that cycleways and pedestrian streets increase
the chance of observing \emph{active} trips in mobile phone records
located near them, whereas highways increase the chance of observing
\emph{motorised} and \emph{taxi}; we assume that \emph{mass-transit} is
associated with primary streets and the rail network as well.

\hypertarget{official-data-sources-from-2020}{%
\subsection{Official Data Sources from
2020}\label{official-data-sources-from-2020}}

We integrated the following official data sources: a socio-economic
characterisation survey (2020), a travel survey (2012), statistics
regarding car permits and global metro usage from the National Institute
of Statistics (2020). The travel survey provides an initial estimation
of the mode split, whereas the other sources are used to update the
initial estimate, jointly with the mobile phone data described earlier.
In addition, we incorporate origin-destination matrices for mass transit
according to smart card data for 2012 and 2020 to validate the model. We
describe these datasets in the section Smart Card Data.

\begin{figure}
\hypertarget{fig:initial_share}{%
\centering
\includegraphics{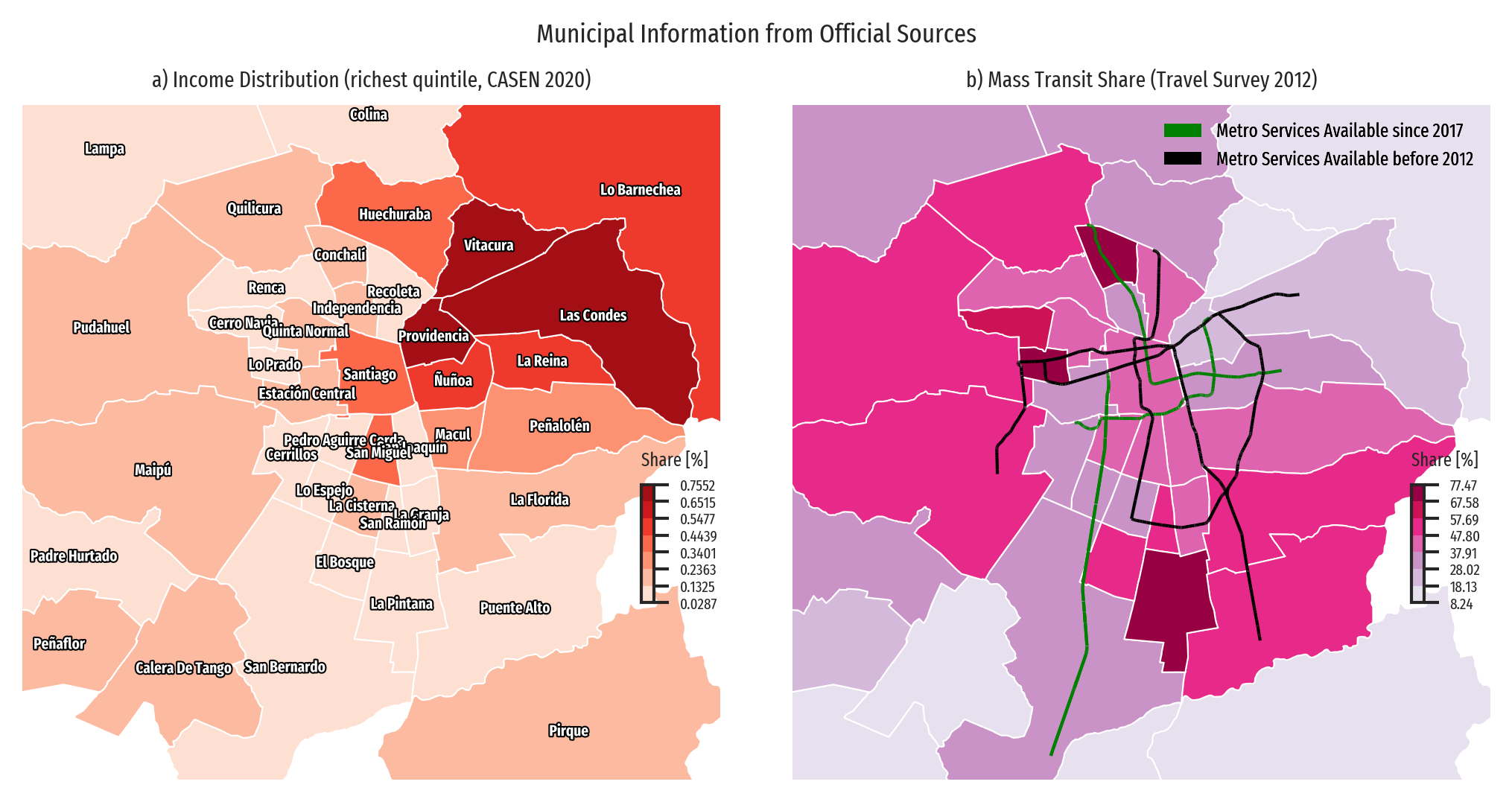}
\caption{Geographical distributions of: a) the richest household income
quintile, from CASEN (2020); b) share of mass transportation in morning
commute time, per municipality, from the Santiago Travel Survey
(2012).}\label{fig:initial_share}
}
\end{figure}

The survey for socio-economic characterisation is named CASEN
(\protect\hyperlink{ref-casen2020}{Ministerio de Desarrollo Social y
Familia, 2020}). Its 2020 edition is the last one released at the time
of reporting this work. Relevant questions for our work include the
income distribution per municipality (in quintiles, see Figure
\ref{fig:initial_share}.a for the distribution of the richest quintile),
the distribution of professions and occupancies per municipality, and
the distribution of migrants with respect to their country of origin per
municipality. These variables are relevant for our methods as Santiago
has experienced migration waves in recent years and a social outburst in
2019. Both have caused changes in work habits, such as an increase in
remote work in areas affected by political protests, and mobility
habits, as migrants have different mobility choices than locals
(\protect\hyperlink{ref-rowe2020drivers}{Rowe and Bell, 2020}).

\begin{figure}
\hypertarget{fig:eod}{%
\centering
\includegraphics{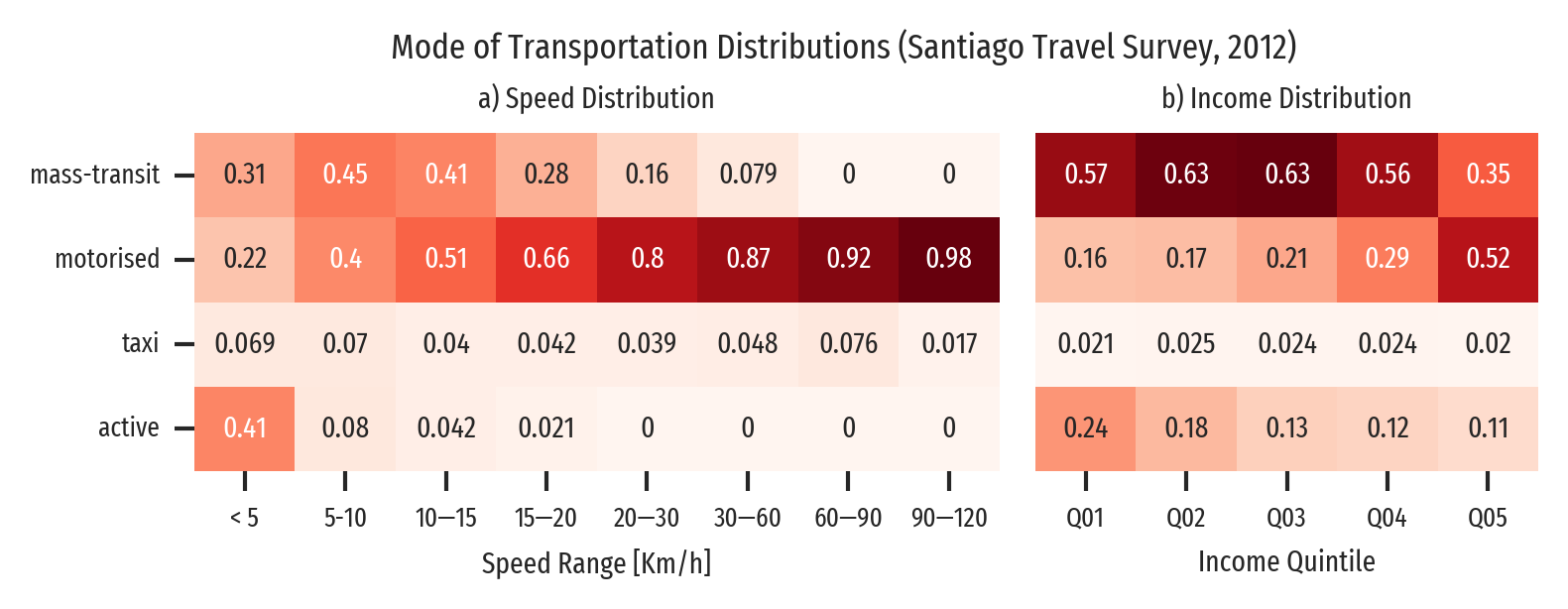}
\caption{Distribution of speed ranges (a) and income quintiles (b) per
mode of transportation (lowest income is Q01, highest income is Q05).
Each matrix is column-normalised.}\label{fig:eod}
}
\end{figure}

The Santiago Travel Survey (EOD, from its name in Spanish) was collected
in 2012 by the Chilean Ministry of Transport and Telecommunications
(\protect\hyperlink{ref-sectra2012informe}{SECTRA, 2015}). As our mobile
phone data were from labour days within a week in March 2020, we
estimated an initial mode split for the municipalities under study in a
workday in morning commute hours (see Figure \ref{fig:initial_share}.b
for the \emph{mass-transit} municipal share, c.f. the income
distribution of Figure \ref{fig:initial_share}.a, which is negatively
correlated). This estimation aggregates all trips originating at each
municipality per mode of transportation, and sums their survey weights.
Following the same procedure, we also estimated the speed range
distribution per mode of transportation (see Figure \ref{fig:eod}.a), as
it can be a relevant feature when discriminating the mode split in OD
flows (\protect\hyperlink{ref-wang2010transportation}{Wang et al.,
2010}); however, these distinctions may be unreliable in high traffic
areas, which is something common at commuting times due to how work
areas are concentrated in Santiago
(\protect\hyperlink{ref-suazo2019displacement}{Suazo-Vecino et al.,
2019}). Additionally, since the travel survey contains income
information, it is also possible to estimate a mode split per income
quintile, that is, the total sum of survey weights of trips per mode
performed by people with a given home income. There is a high
association between motorised transport and higher income levels (see
Figure \ref{fig:eod}.b), and thus, imputed income according to home
municipality could be a relevant signal for mode inference.

The National Institute of Statistics (INE, from its initials in Spanish)
publishes yearly information regarding car permits per municipality,
smart card transactions at the whole Metro (subway) network, and
population composition in 2012 and 2020
(\protect\hyperlink{ref-ine2022}{Instituto Nacional de Estadísticas,
2022}). These statistics inform our method. On the one hand, car permits
must be renewed yearly, and, as such, they are a proxy of how much the
motorised modal share has changed: on average, car permits have
increased by 31.76\% between 2012 and 2020. On the other hand, the
number of smart card transactions at Metro stations has decreased by
43\% from 2012 to 2020, even though there have been new metro and rail
lines since 2012. There are potential explanations for this, including
the effect of the social outburst, where several stations were burned.
As such, the official sources indicate that \emph{motorised} usage has
increased globally, and Metro usage, which is part of
\emph{mass-transit}, has decreased.

We expect these official sources to aid our proposed model when
estimating an updated mode split. Some of these variables may have
importance when estimating the mode split (such as income, car permits,
etc.). Other variables provide demographic controls for the model, which
may help to modulate potential biases in the anonymised data (such as
the population distributions from 2012 and 2020).

\hypertarget{mobile-phone-data}{%
\subsection{Mobile Phone Data}\label{mobile-phone-data}}

\begin{figure}
\hypertarget{fig:xdr}{%
\centering
\includegraphics{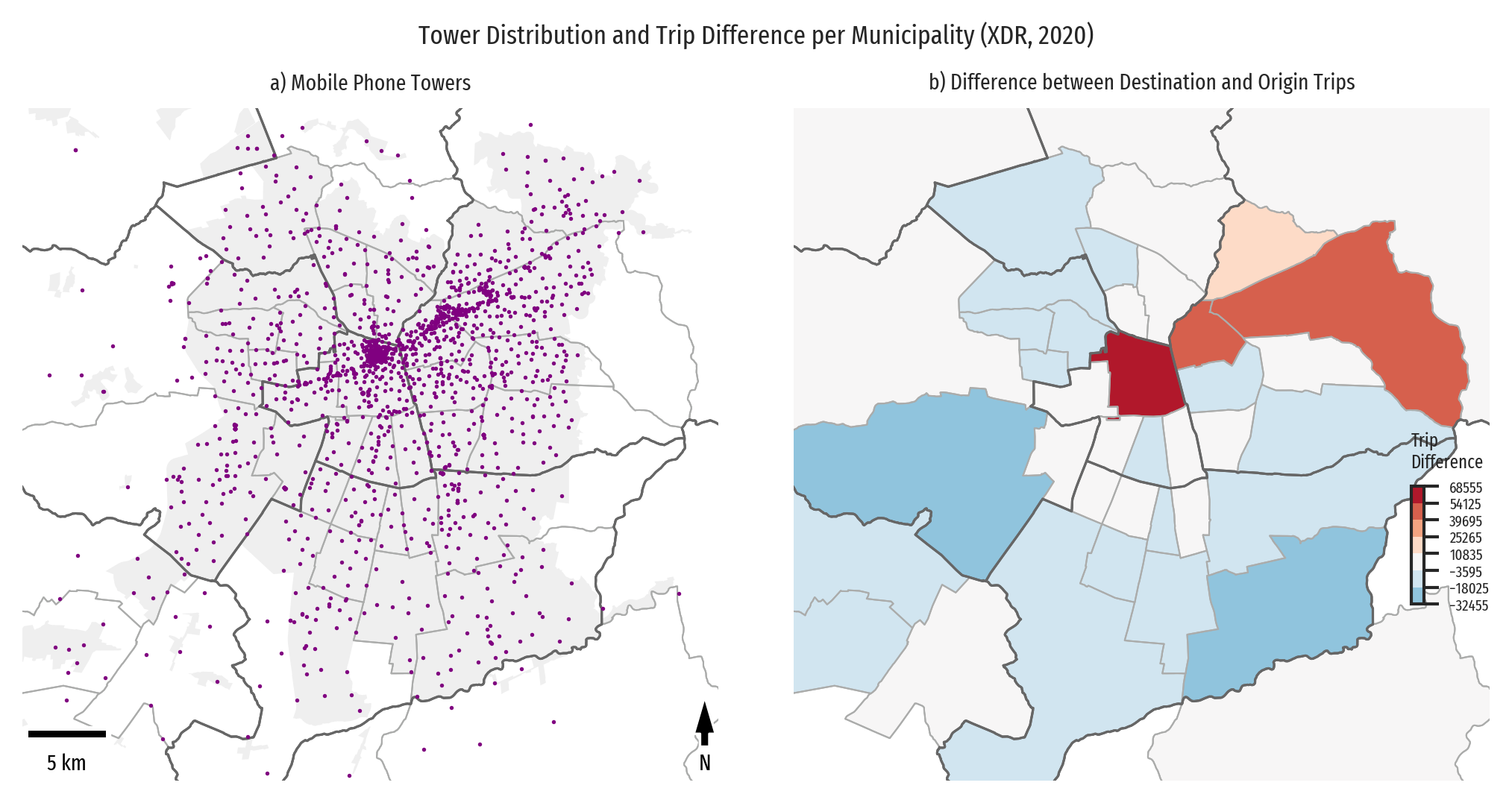}
\caption{a) Distribution of mobile phone towers (also denoted waypoints)
in the area under study. Each dot is a mobile phone tower. b) Each
municipality is coloured according to the difference between inbound and
outbound trips; red municipalities have a positive flux, that is,
attracting more trips than those generated by them, whereas blue
municipalities have a negative flux. Red areas concentrate work and
wealthy areas in the city.}\label{fig:xdr}
}
\end{figure}

We used two mobile phone datasets provided by the telecommunications
operator Telefonica Movistar in Chile: aggregated trajectories from
eXtended Detail Records (XDRs), and aggregated traﬃc from Deep Packet
Inspection (DPI). Both datasets were generated from 1.18M phones active
in the period and area under study. The operator has nearly one-third of
the market in the country. In the urban area of Santiago, it operates
2076 towers distributed in the area of study (see Figure
\ref{fig:xdr}.a). Note that towers are not distributed uniformly in the
city. According to the mobile phone operator, the distribution follows
mobile population patterns in order to ensure quality of service. We
studied municipalities with at least 5 towers in the area under study,
resulting in 40 municipalities, with an average of 48.27 towers (see
Supplementary Material A1 for a characterization of the tower
distribution). As mentioned in the previous data description, we focused
on the morning commute time (from 6 AM to 9 AM), as it is the most
generalisable period that can be studied.

The XDRs dataset is derived from the mobile phone network and is
generated through the polling of device locations
(\protect\hyperlink{ref-huang2019transport}{Huang et al., 2019}). Each
record in the dataset represents a billable event, including calls, SMS,
and Internet downloads. It is important to note that a single record may
encompass multiple events, such as multiple downloads or multiple
connections during a phone call. To determine the most representative
tower for these events, a criteria defined by the company is employed.
For calls, this typically corresponds to the originating call tower,
while for Internet downloads, it is often the most frequently used tower
within a given set of events. Furthermore, not all records are generated
equally, as the company profiles data usage by its users and has
different billing frequencies depending on how active each customer is.
On average, the records are generated in 15-minute intervals for each
device in the Telefónica Movistar network in Santiago
(\protect\hyperlink{ref-graells2016sensing}{Graells-Garrido et al.,
2016}). Each record contains the ID of the representative tower the
device was connected to, a timestamp, and an anonymised device ID.
Device IDs are consistent in the dataset, i.e., two connections with the
same ID describe the trajectory of the same device. We used XDRs to
compute trips using a stop-based method
(\protect\hyperlink{ref-graells2018inferring}{Graells-Garrido et al.,
2018b}). This method compares the timestamp and distance between two
consecutive events. If the speed of the transition is above a threshold
of 0.5 Km/h, the transition encodes part of a trip. This threshold
discards slow transitions that could be associated with movement inside
a venue, for instance, the device of someone who works in a university
will connect to different towers when moving around the venue. Adjacent
parts of a trip are chained to identify its origin and destination
towers, as well as the intermediary towers that serve as waypoints for
each trip. In total, we identified 19M trips in the whole period of
study. Of these, 3M were detected during the morning commute period,
between 6 and 9 AM (Figure \ref{fig:xdr}.b shows the total influx minus
outflux of trips for each municipality at that period).

\begin{figure}
\hypertarget{fig:xdr_trips}{%
\centering
\includegraphics{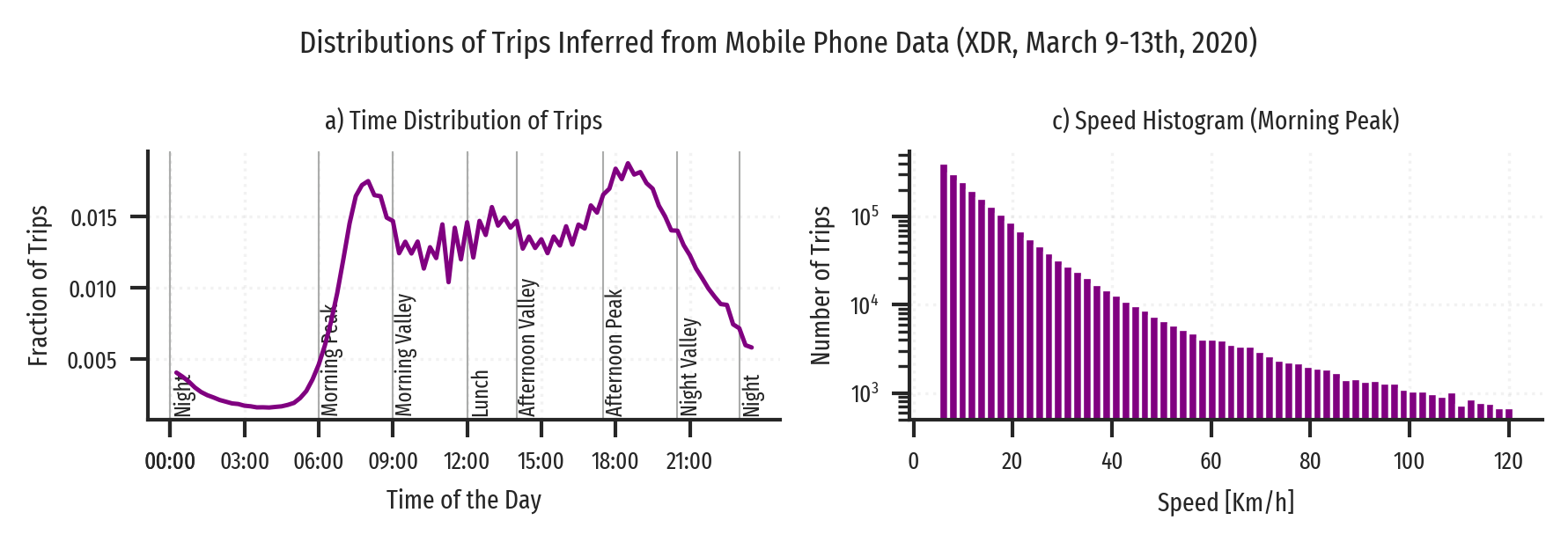}
\caption{a) Distribution of trips observed in mobile phone data
according to the time of the day (purple line). Each line marks the
start of a period of time defined in the Santiago Travel Survey. In this
paper, we focus on commuting times at \emph{morning peak}. b) Histogram
of speeds in observed trips from XDRs.}\label{fig:xdr_trips}
}
\end{figure}

Each trip has features, such as date and time, average speed, and
waypoints. We observe that the trip distribution through all days
resembles a typical city routine with two main peaks in trip volume, one
in morning commute times and another in afternoon commute times (see
Figure \ref{fig:xdr_trips}.a). The distribution of trip speed is skewed
toward lower speeds (see Figure \ref{fig:xdr_trips}.b), which is
expected in an urban setting due to mainly commuting traffic. Note that
we analysed trips with speeds between 5 Km/h, which is just below the
average human speed walking
(\protect\hyperlink{ref-browning2006effects}{Browning et al., 2006}),
and 120 Km/h, which is the legal maximum speed in Chile. Of the 3M
trips, only 7.4K had speeds above 120 Km/h, and 877K had speeds below 5
Km/h. Lastly, waypoints are the intermediary towers that were part of
the trajectory of each trip. These waypoints are important when
inferring modes of transportation, as some waypoints are strongly
associated with each mode, such as towers installed within metro
stations or near highways
(\protect\hyperlink{ref-graells2018inferring}{Graells-Garrido et al.,
2018b}).

The DPI data contain the number of connections in each tower to the 963
most accessed Internet domains from mobile phones (note that, unlike
XDRs, this is not device-level data). These domains can be mapped to
mobile phone applications or Web services, as company or service names
are encoded in domain addresses. The number of domains was determined
after manually removing spyware or other malicious domains, ad-tracking
and other types of analytics performed on apps, removing apps that
showed concentration of usage in the city, that is, they are accessed
from a small set of locations (defined as those in the lower 10\% of
entropy with respect to the number of requests per tower), and unifying
domain names to avoid duplicate app identifiers, for instance, the
domains \emph{api.example.com} and \emph{maps.example.com} are unified
into \emph{example.com}. Thus, whereas XDRs indicates trajectories of
devices, DPI indicates digital activities performed by the people
connected to each tower through mobile phone applications
(\protect\hyperlink{ref-graells2022measuring}{Graells-Garrido et al.,
2022}). Some of these applications may aid the identification of
previous hard-to-detect modes, such as pedestrian, non-motorised and
taxi (due to ride-hailing applications such as Uber and Cabify). Other
apps or activities are important too, as they may be held during
transportation (\protect\hyperlink{ref-jain2008gift}{Jain and Lyons,
2008}). Previous work has found that many types of activities or
applications are relevant to identify mobility patterns
(\protect\hyperlink{ref-graells2022measuring}{Graells-Garrido et al.,
2022}, \protect\hyperlink{ref-graells2018and}{2018a}), as such, we
analyse a wide spectrum of apps, without focusing on transportation apps
only. For instance, they may be related to the mode of transportation of
the device owner (\protect\hyperlink{ref-graells2018and}{Graells-Garrido
et al., 2018a}). From a total of 1278 non-duplicate apps (or domain
names), we kept 963 for analysis after filtering using the entropy
criteria.

\begin{figure}
\hypertarget{fig:dpi}{%
\centering
\includegraphics{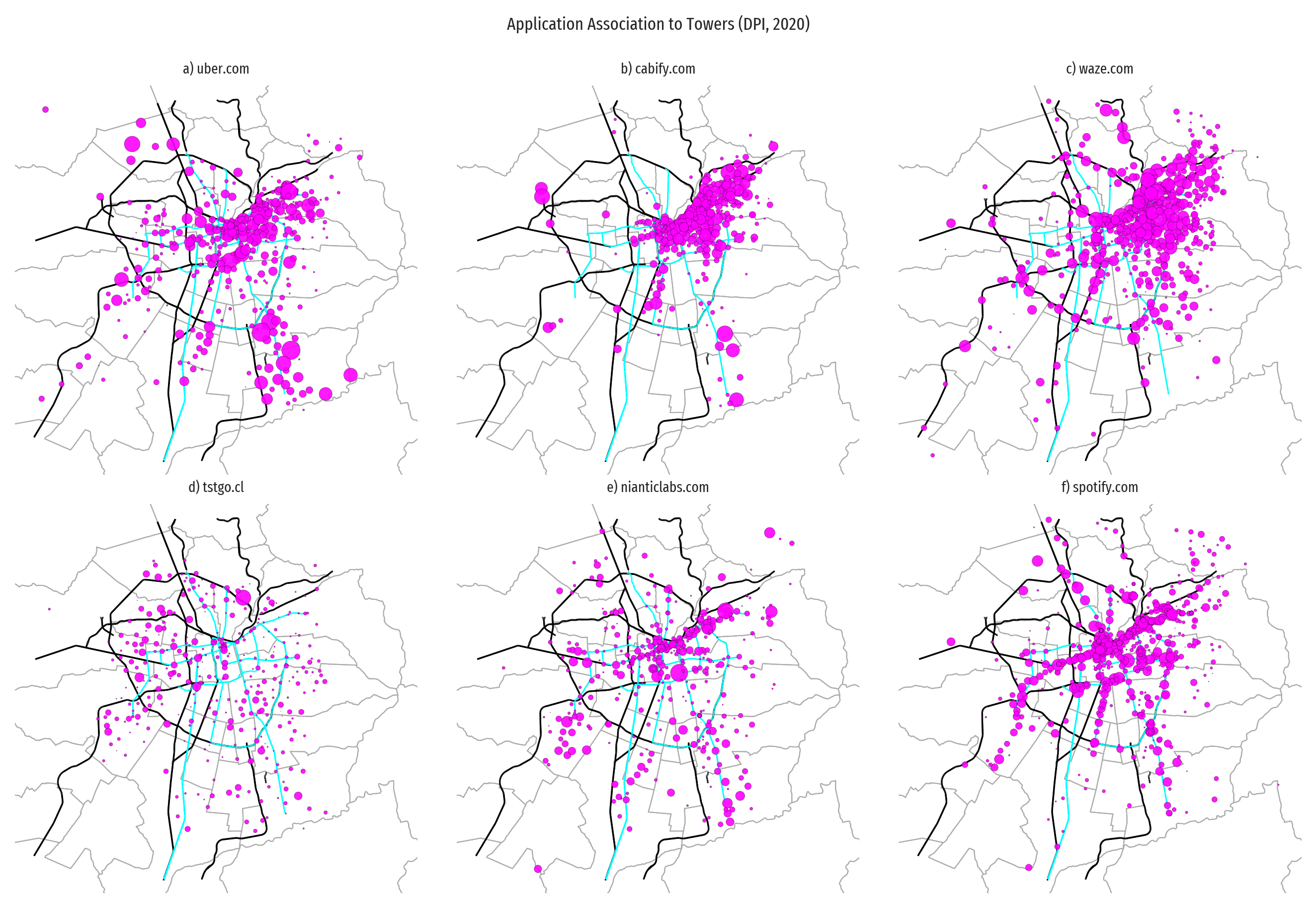}
\caption{Distribution of mobile application usage (from DPI data) in
towers. Each tower is depicted with a circle with area proportional to
the intensity of the application usage. a) Uber (ride-hailing). b)
Cabify (ride-hailing). c) Waze (GPS routing). d) Transantiago
(mass-transit). e) Niantic Labs (augmented reality games such as Pokémon
Go). e) Spotify (audio). Black lines represent primary streets in the
city, cyan lines represent the metro network. Urban network data has
copyright from OpenStreetMap contributors, used with
permission.}\label{fig:dpi}
}
\end{figure}

As app usage may be related to the mode split by associating waypoints
(as seen in XDRs) with activities (as seen in DPI), we estimated the
distribution of mobile phone applications accessed at each waypoint
(tower). This distribution can be expressed as a matrix (see next
section for details of this and other matrices) where each cell contains
the total access count to an app reported in the data at a given tower.
We applied the Log-odds ratio with Uninformative Dirichlet Prior formula
to the access counts per application
(\protect\hyperlink{ref-monroe2008fightin}{Monroe et al., 2008}). This
formula considers the frequency and variability of access to each
application, and determines which waypoints had relevant values over a
Dirichlet prior. Figure \ref{fig:dpi} displays the association of each
tower at commuting times with example applications after applying this
transformation: a) Uber, a ride-hailing app related to \emph{taxi}; b)
Cabify, another ride-hailing app, related to \emph{taxi}; c) Waze, a GPS
routing, related to \emph{motorised} and \emph{taxi}; d) Transantiago,
the official domain of the \emph{mass-transit} system; e) Niantic Labs,
the creator of augmented reality games such as Pokémon Go, which could
be related to \emph{active} transport
(\protect\hyperlink{ref-graells2017effect}{Graells-Garrido et al.,
2017}); and e) Spotify, an audio app with a distribution that resembles
the metro network (\emph{mass-transit}).

Thus, with mobile phone data, there are observations of trips, with
their corresponding speeds and waypoints, and information associated
with those waypoints. To the extent of our knowledge, this combination
of XDRs and DPI has not been used in modal split problems before. Then,
we expect to obtain an updated modal split by integrating these data
with the official data previously described.

\hypertarget{smart-card-data}{%
\subsection{Smart Card Data}\label{smart-card-data}}

Finally, we include an additional dataset from the system ADATRAP
(\protect\hyperlink{ref-Gschwender2016vi}{Gschwender et al., 2016})
provided by the Direction of Metropolitan Public Transport (DTPM from
its initials in Spanish) for validation purposes. ADATRAP integrates
multiple data sources from the mass transit system in Santiago,
including smart card data, bus GPS data, vehicle operator data, among
others. In Santiago, passengers must validate their \emph{bip!} cards
when boarding a bus or entering a Metro station, but no alighting
validation is required, thus, smart card data only registers the
boarding location of trips. the methodology relied on usage patterns
such as the regularity of commuting trips to estimate the number of
trips made. Hence, a methodology based on recurrent patterns of usage
per user and geometry of distances toward boarding sites is used
(\protect\hyperlink{ref-munizaga2012estimation}{Munizaga and Palma,
2012}). The ADATRAP methodology also includes fare evasion correction
(\protect\hyperlink{ref-munizaga2020fare}{Munizaga et al., 2020}). Trips
are reported at municipality and macro-area levels (a macro-area
contains multiple municipalities).

\begin{figure}
\hypertarget{fig:adatrap_trips}{%
\centering
\includegraphics{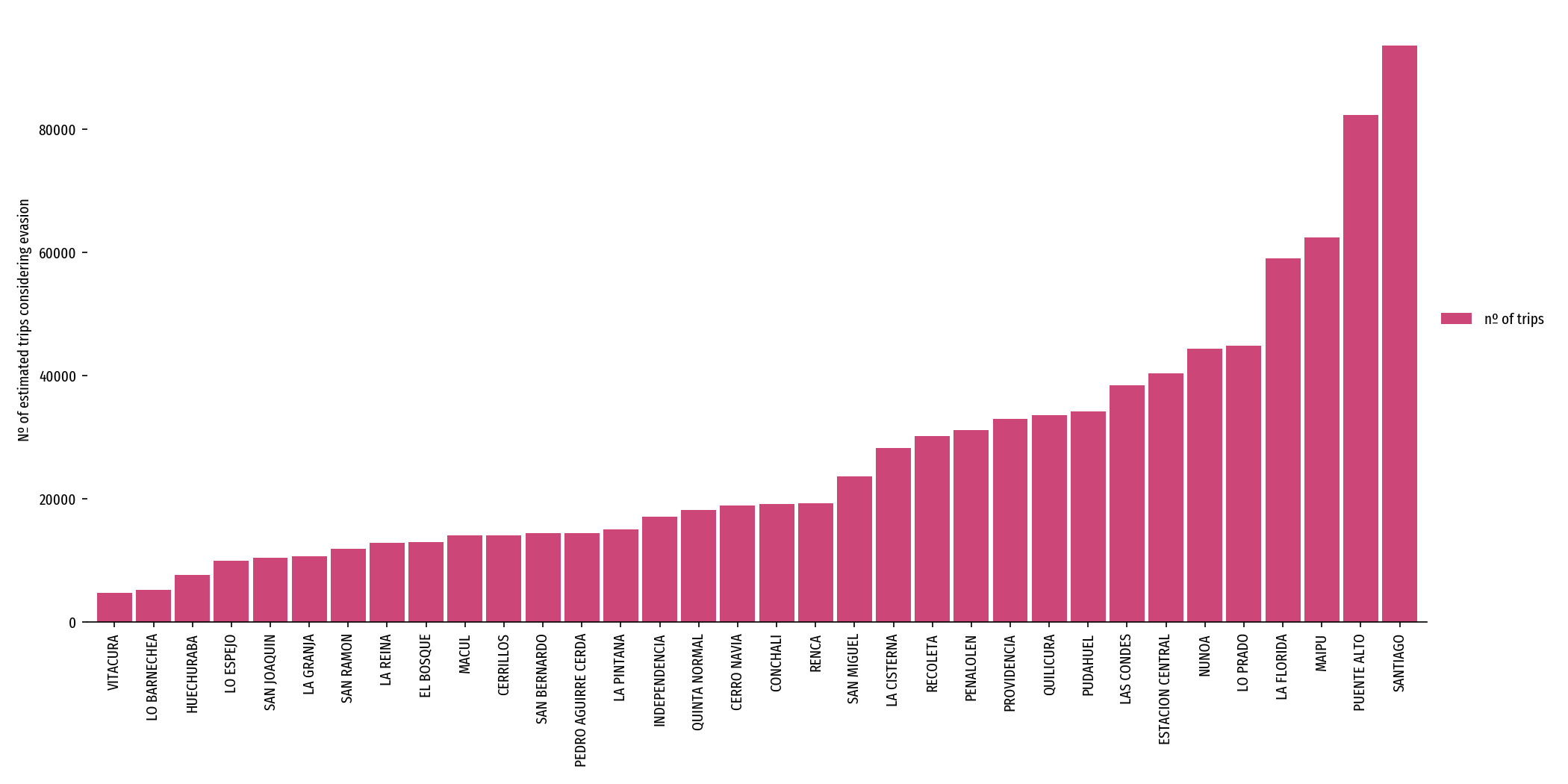}
\caption{Distribution of trips per municipality in smart card data from
ADATRAP.}\label{fig:adatrap_trips}
}
\end{figure}

To ensure a rigorous validation of our results, ADATRAP estimated
\emph{mass-transit} trip counts from commuting times using smart card
data between March 9, 2020 to March 13, 2020, from buses and metro .
Specifically, we focused on trips during the morning peak period (06:00
AM - 09:00 AM). Given that ADATRAP provides trip counts with the
corresponding time of day and date, we calculated the average daily trip
count during the commuting times for the specified dates and study
period. In total, the system reported 932K trips in mass transit during
the morning peak period in Santiago for the analysed dates.

\hypertarget{methods}{%
\section{Methods}\label{methods}}

\begin{figure}
\hypertarget{fig:schema}{%
\centering
\includegraphics{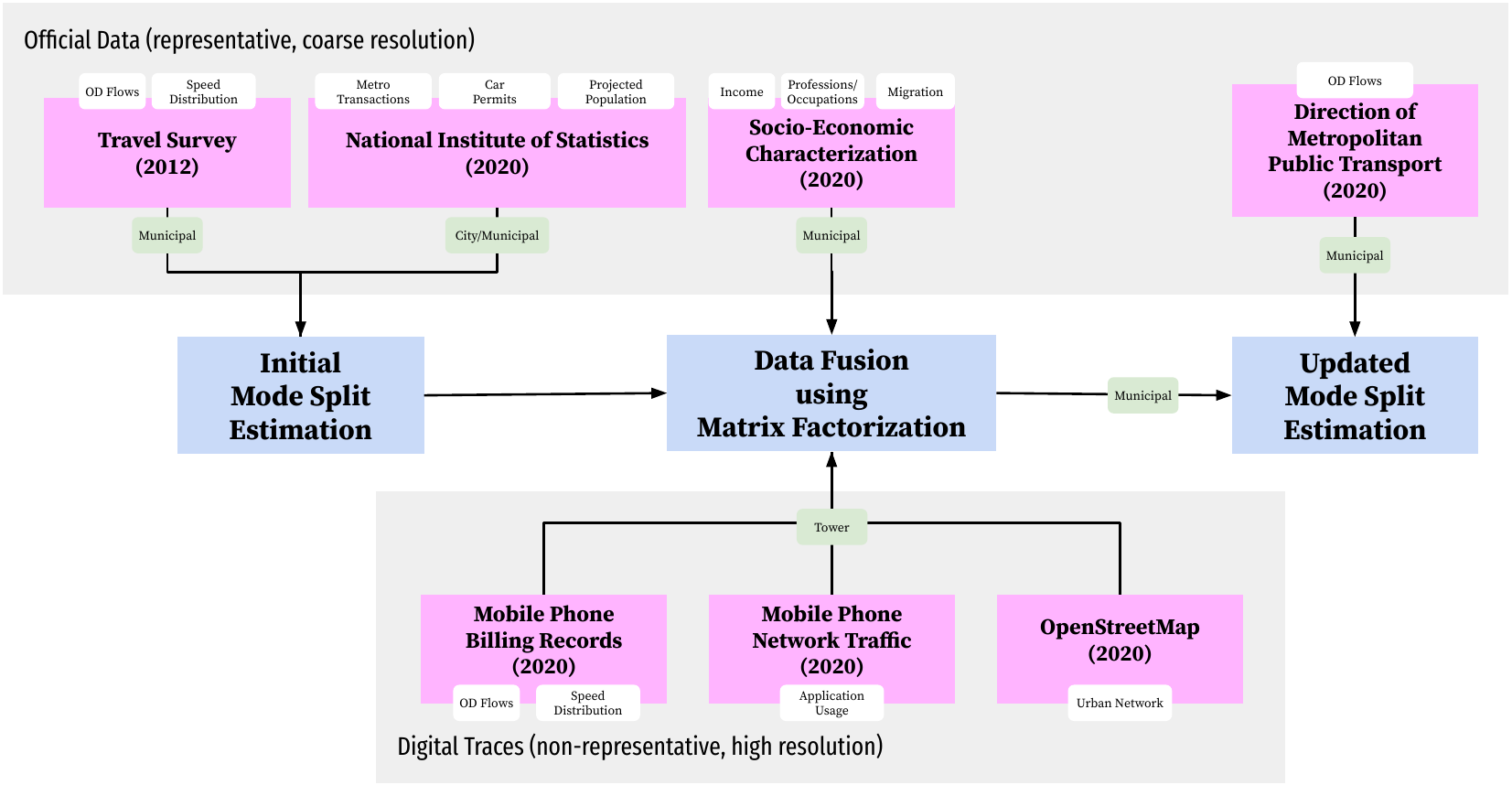}
\caption{Schematic view of the methods and data used to estimate an
updated mode split for a city.}\label{fig:schema}
}
\end{figure}

Here we propose a method that processes traditional and digital data
sources (see an overview of the method in Figure \ref{fig:schema}). The
method consists of three main stages:

\begin{enumerate}
\def\labelenumi{\arabic{enumi})}
\item
  \emph{Initial mode split estimation}, where official sources and
  domain \textgreater{} knowledge are used to build a candidate mode
  split for all \textgreater{} administrative units.
\item
  \emph{Data fusion using matrix factorization}, a procedure that
  generates \textgreater{} a latent representation of several datasets
  represented in an \textgreater{} unified manner. As data we consider
  official and digital sources, \textgreater{} with features related to
  mobility patterns, socio-demographic \textgreater{} characteristics,
  and urban infrastructure.
\item
  \emph{Updated mode split estimation}, where generate an updated mode
  \textgreater{} split that takes into account all available information
  in the \textgreater{} latent representations from the previous point.
\end{enumerate}

As a case study, we work with data from Santiago, Chile. The
administrative units under study are the several municipalities in the
Santiago Metropolitan Area. We used the data sources described in the
previous section. Next, we describe the pre-processing and methods we
apply to these data.

\hypertarget{initial-guess-of-the-updated-mode-split}{%
\subsection{Initial Guess of the Updated Mode
Split}\label{initial-guess-of-the-updated-mode-split}}

The first step in our methodology is to build an initial guess of an
updated mode split for the city. Recall that there are observations from
official data that can aid an initial estimation of an updated mode
split, and we can use expert knowledge of the city as well. On the one
hand, the National Institute of Statistics (INE) provides a city-wide
count of smart card transactions in the Metro system and the number of
car permits per municipality, as well as population projections. These
statistics are available for 2012 and 2020. On the other hand, we assume
that \emph{taxi} trips have increased, due to the emergence of
ride-hailing applications, although this increase is modest at commuting
times (\protect\hyperlink{ref-tirachini2019ride}{Tirachini and Río,
2019}); we also assume that pedestrian trips have decreased due to the
effects of the social outburst from October 2019. Taking these factors
into account, and starting from the trip counts per mode of
transportation \(m\) as measured by the Santiago Travel Survey 2012, we
use the updated population change from 2012 to 2020 and the
aforementioned factors to estimate an initial guess of the updated mode
split in the following way:

\begin{eqnarray}
MT_{20}(m) & =& MT_{12}(m) \times \frac{Population_{20}(m)}{Population_{12}(m)} \times \sqrt{\frac{Metro_{20}}{Metro_{12}}} \\
C_{20}(m) & = & C_{12}(m) \times \frac{Population_{20}(m)}{Population_{12}(m)} \times \sqrt{\frac{Permits_{20}(m)}{Permits_{12}(m)}} \\
A_{20}(m) & = & A_{12}(m) \times \frac{Population_{20}(m)}{Population_{12}(m)} \times 0.975 \\
T_{20}(m) & = & (T_{12}(m) + 1) \times \frac{Population_{20}(m)}{Population_{12}(m)} \times 1.09. \\
\end{eqnarray}

In these equations, \emph{m} is a municipality, \emph{Population} is the
corresponding municipal population at the years 2012 and 2020,
\emph{Metro} is the number of smart card transactions, \emph{Permits} is
the corresponding municipal car permits at the years 2012 and 2020;
\emph{MT} is \emph{mass-transit}, \emph{C} is \emph{motorised}
transport, \emph{A} is \emph{active} transport, and \emph{T} is
\emph{taxi}. In all modes of transportation we multiply the number of
trips per municipality \emph{m} with their ratio of change according to
the population distribution. The number of trips is the daily average
for the period of study for 2020, to make it comparable to the trip
count obtained from the travel survey (i.e., an average commuting trip
count for laboral days in non-estival periods).

We adjusted \emph{MT} according to available data: metro trips were
reduced by 43.70\%, however, public transport includes buses and trains,
and the metro network was severely affected by the social outburst from
2019, with many stations closed or not operative during the period under
study. Hence, we modulate the change in smart card records using a
square root. For motorised trips, we adjusted \emph{C} using a similar
approach. In addition to updating the amount of trips according to
population change, we modulate them using the change in car permits per
municipality. We also modulate this rate of change with a square root,
because not all cars are for commuting, as others are for secondary
usage at each household, and others are for work. Hence, we modulate the
car permit for those situations. Active transport \emph{A} follows a
similar schema. Here, we assume that those trips have reduced their
relative share by 2.5\%, given the social context of the city. And
finally, for taxi trips \emph{T} we applied the population change factor
and the increase factor based on previous reports on ride-hailing usage
(\protect\hyperlink{ref-tirachini2019ride}{Tirachini and Río, 2019}).
Note that we added one trip to the observed taxi trip count in 2012 to
avoid having zeroes in the updated modal partition, as some
municipalities did not have taxi trips in 2012 at commuting times in
laboral days.

The ratio between this initial mode split estimation and a naive one
where the original R01 is multiplied municipality-wise with the fraction
of population change is 0.9634. These equations are fairly direct and
new data sources or modulation factors can be introduced according to
expert knowledge if needed.

\hypertarget{data-integration-through-an-unified-representation}{%
\subsection{Data Integration through an Unified
Representation}\label{data-integration-through-an-unified-representation}}

The datasets described earlier were generated by different sources,
although they contain several shared concepts. For instance, all of them
have data related to the municipalities in the area under study, either
explicitly, or through aggregation of data. Here we describe a unified
representation that enables data integration.

First, we identified all different concepts present in the data (e.g.,
municipality, mode of transportation, waypoint, etc.; see Table 1 for
all of them, including description). Under this definition, every table
from the datasets expresses a relation between concepts as a matrix. For
instance, the mode split for different speed ranges (see Figure
\ref{fig:eod}.a) is a matrix, where rows are modes of transportation,
columns are speed ranges, and cell values quantify the fraction of trips
that correspond to the specific mode/speed range pair. Analogously, the
mode split per municipality is also a matrix.

\clearpage

\begin{longtable}[]{@{}
  >{\raggedright\arraybackslash}p{(\columnwidth - 8\tabcolsep) * \real{0.1954}}
  >{\raggedright\arraybackslash}p{(\columnwidth - 8\tabcolsep) * \real{0.2529}}
  >{\raggedright\arraybackslash}p{(\columnwidth - 8\tabcolsep) * \real{0.1609}}
  >{\raggedright\arraybackslash}p{(\columnwidth - 8\tabcolsep) * \real{0.1609}}
  >{\raggedright\arraybackslash}p{(\columnwidth - 8\tabcolsep) * \real{0.2069}}@{}}
\caption{List of concepts expressed in our datasets.}\tabularnewline
\toprule()
\begin{minipage}[b]{\linewidth}\raggedright
Concept
\end{minipage} & \begin{minipage}[b]{\linewidth}\raggedright
Description
\end{minipage} & \begin{minipage}[b]{\linewidth}\raggedright
Datasets
\end{minipage} & \begin{minipage}[b]{\linewidth}\raggedright
Cardinality
\end{minipage} & \begin{minipage}[b]{\linewidth}\raggedright
Example Values
\end{minipage} \\
\midrule()
\endfirsthead
\toprule()
\begin{minipage}[b]{\linewidth}\raggedright
Concept
\end{minipage} & \begin{minipage}[b]{\linewidth}\raggedright
Description
\end{minipage} & \begin{minipage}[b]{\linewidth}\raggedright
Datasets
\end{minipage} & \begin{minipage}[b]{\linewidth}\raggedright
Cardinality
\end{minipage} & \begin{minipage}[b]{\linewidth}\raggedright
Example Values
\end{minipage} \\
\midrule()
\endhead
Application & Application or web domain accessed by mobile phones. & DPI
(2020) & 963 & uber.com, transantiago.cl \\
Income & Mean income at a given area, in quintiles. & EOD (2012), CASEN
(2020) & 5 & Q01, Q05 \\
Migration & Country of origin of migrant population. & CASEN (2020) & 55
& Spain, Venezuela \\
Mode of Transportation & Type of vehicle used for commuting trips. & EOD
(2012) & 4 & mass-transit, motorised \\
Municipality & Administrative unit used for analysis & All & 40 &
Santiago, Providencia \\
Population & Demographic characteristics (age in decades) per year & INE
(2012, 2020) & 22 & 0 to 10 years, \textgreater{} 65 years \\
Urban Infrastructure & Characteristics of the built environment related
to transportation & OSM (2020) & 7 & near railways, near highways \\
Speed & Average speed, aggregated in ranges & EOD (2012), XDR (2020) & 8
& \textless{} 5 Km/h, 30 to 60 Km/h \\
Waypoint & A mobile phone tower & DPI (2020), XDR (2020) & 1878 &
BTS-0001, BTS-0087 \\
Work Type & A profession or occupation & CASEN (2020) & 10 & scientist,
machine operator \\
\bottomrule()
\end{longtable}

\begin{figure}
\hypertarget{fig:network}{%
\centering
\includegraphics{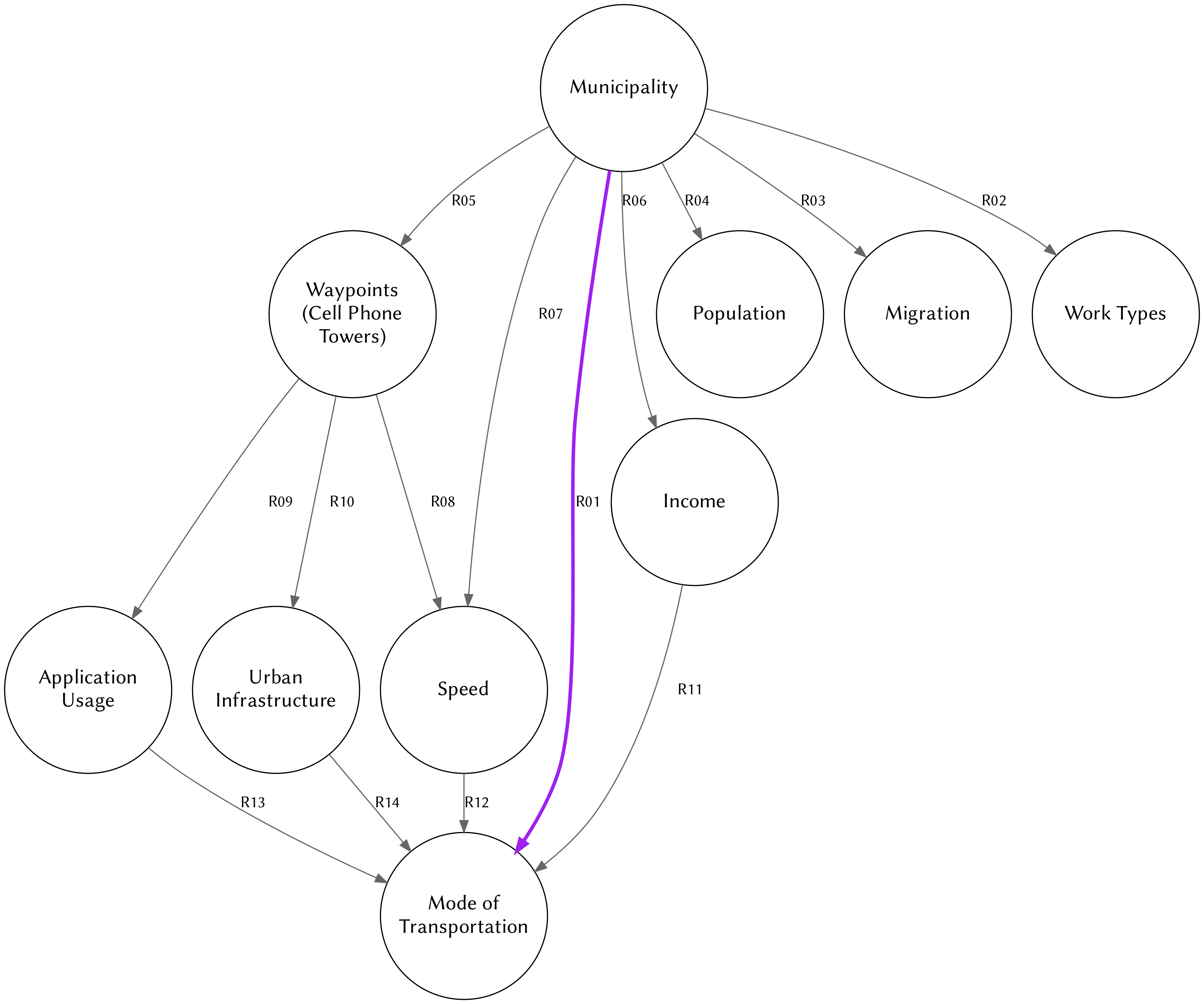}
\caption{Diagram of concept relations from our datasets. Nodes (circles)
are the different concepts we analyse from the data. Each edge is a
relation between concepts, which we formally represent as a matrix from
the dataset. The purple arrow represents the mode split per
municipality.}\label{fig:network}
}
\end{figure}

Second, we built a relationship network (see Figure \ref{fig:network})
with all matrices under analysis (see Table 2). We work with a total of
14 matrices referred to as \emph{Rn}, where n ranges from 1 to 14. Most
of these matrices were built directly from the data. Note, however, that
two of these matrices, Application Usage and Mode of Transportation
(R13), and Urban Infrastructure and Mode of Transportation (R14) were
manually built by us. For example, R13 was built relating apps domains
such as \emph{uber.com} and \emph{transantiago.cl}, with modes of
transportation such as taxi and mass-transit . We explored the full list
of apps/domains to establish these associations and found 53 apps that
could be associated with at least one mode of transportation. Some
domains are associated with two. For instance, waze.com may be used by
any \emph{motorised} vehicle, and this includes the \emph{taxi}
category. Additionally, the Municipality and Waypoints matrix (R05) was
weighted using TF-IDF, a common weighting method that normalises each
row in the matrix. Previous experiments have shown that such weighting
increases the stability and accuracy of the model
(\protect\hyperlink{ref-graells2018inferring}{Graells-Garrido et al.,
2018b}). Supplementary Material Section A2 shows a visualisation of all
relations in the dataset.

The association between domains and modes of transport goes from 0 (no
assoc.) to 1 (full assoc.). This leaves 1,007 apps without association.
We assign a value of 0.25 to these (so all associations for an app sum
1). Similarly, for the R14 matrix, the matrix which relate the urban
infrastructure of the city to a mode of transport, was manually assigned
using a boolean flag (0 or 1) to each pair made of infrastructure type
and mode of transportation, as described in the OpenStreetMap section
earlier.

The R01 matrix contains the initial solution of the mode split per
municipality estimated earlier in this section. In the remainder of this
paper we describe a method to obtain an updated R01, as the other
matrices contain information that may help in improving the updated mode
split.

\clearpage
\begin{longtable}[]{@{}
  >{\raggedright\arraybackslash}p{(\columnwidth - 8\tabcolsep) * \real{0.0682}}
  >{\raggedright\arraybackslash}p{(\columnwidth - 8\tabcolsep) * \real{0.1932}}
  >{\raggedright\arraybackslash}p{(\columnwidth - 8\tabcolsep) * \real{0.1932}}
  >{\raggedright\arraybackslash}p{(\columnwidth - 8\tabcolsep) * \real{0.3750}}
  >{\raggedright\arraybackslash}p{(\columnwidth - 8\tabcolsep) * \real{0.1477}}@{}}
\caption{All relations analysed. Each relation is a matrix built
fromvthe data.}\tabularnewline
\toprule()
\begin{minipage}[b]{\linewidth}\raggedright
ID
\end{minipage} & \begin{minipage}[b]{\linewidth}\raggedright
Source Concept
\end{minipage} & \begin{minipage}[b]{\linewidth}\raggedright
Target Concept
\end{minipage} & \begin{minipage}[b]{\linewidth}\raggedright
Cell Value Description
\end{minipage} & \begin{minipage}[b]{\linewidth}\raggedright
Source
\end{minipage} \\
\midrule()
\endfirsthead
\toprule()
\begin{minipage}[b]{\linewidth}\raggedright
ID
\end{minipage} & \begin{minipage}[b]{\linewidth}\raggedright
Source Concept
\end{minipage} & \begin{minipage}[b]{\linewidth}\raggedright
Target Concept
\end{minipage} & \begin{minipage}[b]{\linewidth}\raggedright
Cell Value Description
\end{minipage} & \begin{minipage}[b]{\linewidth}\raggedright
Source
\end{minipage} \\
\midrule()
\endhead
R01 & Municipality & Mode of Transportation & The number of trips
originated at a given municipality for a given mode of transportation. &
Initial Estimation using EOD (2012) and INE (2020) \\
R02 & Municipality & Work Types & The number of people in a given
profession that lives in a given municipality. & CASEN (2020) \\
R03 & Municipality & Migration & The number of migrants from a given
origin country that live in a given municipality. & CASEN (2020) \\
R04 & Municipality & Population & The number of people in a given age
range that lives in a given municipality, for two different years. & INE
(2012, 2020) \\
R05 & Municipality & Waypoints & A weighted value estimated from the
number of trips from people that live in a given municipality that
passes through a given waypoint or mobile phone tower. & XDR (2020) \\
R06 & Municipality & Income & The number of people in a given income
quintile that lives in a given municipality. & CASEN (2020) \\
R07 & Municipality & Speed & The number of trips originating at a given
municipality that have an average speed within a given speed range. &
XDR (2020) \\
R08 & Waypoints & Speed & The number of trips that pass through a given
waypoint that have an average speed within a given speed range. & XDR
(2020) \\
R09 & Waypoints & Application Usage & A weighted value estimated from
the number of accesses to a given app/domain from a given waypoint. &
DPI (2020) \\
R10 & Waypoints & Urban Infrastructure & Whether (boolean) there is a
given type of urban infrastructure in a ratio of 500 metres around a
given waypoint. & XDR(2020) and OSM (2020) \\
R11 & Income & Mode of Transportation & The fraction of trips in a given
mode of transportation for a given income quintile. & EOD (2012) \\
R12 & Speed & Mode of Transportation & The fraction of trips in a given
mode of transportation for a given speed range. & EOD (2012) \\
R13 & Application Usage & Mode of Transportation & The association
(between 0 and 1) of a given app/domain with a given mode of
transportation. & Manually built \\
R14 & Urban Infrastructure & Mode of Transportation & Whether (boolean)
trips from a given mode of transportation use a given urban
infrastructure. & Manually built \\
\bottomrule()
\end{longtable}

\hypertarget{data-fusion-using-matrix-factorisation}{%
\subsection{Data Fusion using Matrix
Factorisation}\label{data-fusion-using-matrix-factorisation}}

At this point, we have a set of relationships between concepts. Although
these relationships have a common representation, as all of them are
matrices, they have not been unified yet; in other words, we do not get
new insights from all these relationships being analysed together. In
our context, we would expect to improve our initial guess of mode split
by employing the other relations expressed in the data. Our goal is to
compute a new R01 matrix (mode split) that contains an updated modal
split by using the information present in the rest of the dataset, which
is up-to-date. We do so through Data Fusion, a family of methods to
integrate datasets into an unified representation with the aim of
extracting knowledge that cannot emerge from each data source alone
(\protect\hyperlink{ref-el2011data}{El Faouzi et al., 2011}).

In previous work, matrix factorisation has been used to infer mode of
transportation usage
(\protect\hyperlink{ref-graells2018inferring}{Graells-Garrido et al.,
2018b}). One of the characteristics of matrix factorisation methods is
the ability to reconstruct matrices by means of matrix projection. A
matrix \(R01'\) could be expressed as a reconstruction of \(R01\) based
on the other matrices available on the dataset. Previous work using
factorisation has employed Non-negative Matrix Factorisation (NMF),
which decomposes a positive matrix \(M_{ij}\) into the multiplication of
two matrices: \begin{equation}
\mathbf{M_{ij}} \approx \mathbf{G_i} \times \mathbf{G_j},
\end{equation} where \(G_i\) is a low-rank latent representation of
concept \(i\) (rows of \(M_{ij}\)), and \(G_j\) is a low-rank latent
representation of concept \(j\) (columns of \(M_{ij}\)). In this schema,
if the matrix to be decomposed is \(R01\), then the latent dimensions
are interpreted as modes of transportation
(\protect\hyperlink{ref-graells2018inferring}{Graells-Garrido et al.,
2018b}).

Although NMF is widely used in multiple areas, it is insufficient for
our problem, because there is no direct way of incorporating the
additional information available in the solution. In other words, both
concepts are clustered in the latent space without considering the
network of relations. In this situation, the Matrix Tri-factorization
approach provides a definition of a co-clustering that considers these
connections (\protect\hyperlink{ref-ding2006tnmf}{Ding et al., 2006};
\protect\hyperlink{ref-vzitnik2014data}{Žitnik and Zupan, 2014}), by
separating the latent representation of each concept from the latent
representation of their relationship: \begin{equation}
\mathbf{M_{ij}} \approx \mathbf{G_{i}} \times \mathbf{S_{ij}} \times \mathbf{G_{j}^T},
\end{equation} where \(S_{ij}\) is denoted as the backbone matrix that
encodes the relation \(M_{ij}\) in the latent space, and \(G_i\) (and,
by analogy, \(G_j\)) is the latent representation of the concept \(i\)
in all its relationships. In other words, the G matrices cluster
concepts into latent dimensions, and the S matrices encode the
relationships between clusters of different concepts. The concept
matrices have positive elements only (as in NMF), whereas the backbone
matrices may have negative elements. In this way, the backbone matrix
may express positive and negative associations between clusters of
concepts.

Following this schema, the fully updated mode split matrix \(R01'\) can
be reconstructed from its corresponding backbone matrix \(S\), and the
concept matrices \(G_{mode}\) and \(G_{municipality}\), which should be
computed taking into account the full network of relationships.

To perform the factorisation, the following problem must be solved:
\begin{equation}
\min_{\mathbf{G} >= 0} \sum_{\mathbf{M_{ij}} \in \mathcal{R}} \left\| \mathbf{M_{ij}}-\mathbf{G_i S_{ij} G_j}^{T} \right\|.
\end{equation} Here, \(mathcal{R}\) represents the set of relations in
our data, \(mathbf{M_{ij}}\) denotes a relation matrix between concepts
\(i\) and \(j\), and \(\left\| \cdot \right\|\) represents the Frobenius
norm. Although matrix factorisation is an NP-HARD problem, an efficient
implementation to solve this problem, based on multiplicative updates
(\protect\hyperlink{ref-vzitnik2014data}{Žitnik and Zupan, 2014}), is
available in the scikit-fusion library
(\protect\hyperlink{ref-vcopar2019fast}{Čopar et al., 2019}).

As in other factorisation methods, each concept matrix \(G_i\) requires
a rank parameter \(k_i\). The rank \(k\) determines the dimension of the
latent space that represents the data; one interpretation of its meaning
is how much the data is compressed in its latent representation. For
instance, if \(k\) is very small in comparison to the total dimension of
the concept, there is a risk of information loss due to extreme
compression; conversely, for large values of \(k\), the model may
overfit and give more importance to noise than to the actual information
in the model.

In our context, we aim at compressing the data enough to learn patterns
behind each concept, while at the same time allowing the reconstruction
of the data to be different from its initial value, as we assume that we
are updating the data, rather than merely compressing/decompressing it.
Furthermore, we need to balance the complexity of the model (higher
ranks imply a more complex model) with the computational cost of
performing the factorization. As such, we evaluated several rank
functions and followed the commonly used elbow method to select a
function (see Supplementary Material Section A3 for details). We
selected the following heuristic to estimate the rank \(k_i\) for a
concept \(c_i\): \begin{equation}
k_i = 2 \times \sqrt{\mid c_i \mid} - 1,
\end{equation} Where \(\mid c_i \mid\) is the cardinality of \(c_i\)
(see Table 1).

By solving the optimization problem, all relations in the dataset can be
reconstructed by multiplying the corresponding concept and backbone
matrices. In particular, the reconstructed mode split matrix \(R01'\)
can be obtained directly through the following multiplication:
\begin{equation}
\mathbf{R01'} = \mathbf{G_{municipality}} \times \mathbf{S_{municipality,mode}} \times \mathbf{G_{mode}^T}.
\end{equation} This \(R01'\) matrix is the ultimate result of this
pipeline, as it expresses the modal split in terms of the knowledge
available in the entire dataset, in addition to the initial estimation.

The solution to the optimization problem is guaranteed to be a local
optimum only, due to the method being based on multiplicative updates.
Since this method employs random numbers to initialise and then improve
a solution, the procedure to select the best model is to run multiple
instances, each with a different random seed for initialisation. The
criteria for selection is based on reconstruction error
(\protect\hyperlink{ref-vzitnik2014data}{Žitnik and Zupan, 2014}). Here,
we define the global error \(e\) as the geometric mean of normalised
reconstruction error for all relations, defined as follows:
\begin{equation}
e = \sqrt[|\mathcal{R}|]{\Pi_{\mathbf{M_{ij}} \in \mathcal{R}} \frac{\left\| \mathbf{M_{ij}}-\mathbf{G_i S_{ij} G_j}^{T} \right\|}{\left\| \mathbf{M_{ij}} \right\|}}.
\end{equation} The model with the smallest reconstruction error \(e\)
will be selected. Note that the \(R01'\) matrix is excluded from this
calculation as its difference with the original \(R01\) matrix is not
interpreted as an error, rather, we interpret it as the change in modal
split.

Next, we analyse the results of applying these methods to the data from
Santiago, Chile.

\hypertarget{results}{%
\section{Results}\label{results}}

In this section we report the results of following the data fusion
approach with data from Santiago, Chile. First, we focus on the model
selection procedure. Then, we analyse the updated mode split
reconstructed by the best model. And finally, we validate our results
with an official data source.

\hypertarget{model-selection}{%
\subsection{Model Selection}\label{model-selection}}

\begin{figure}
\hypertarget{fig:model_variability}{%
\centering
\includegraphics{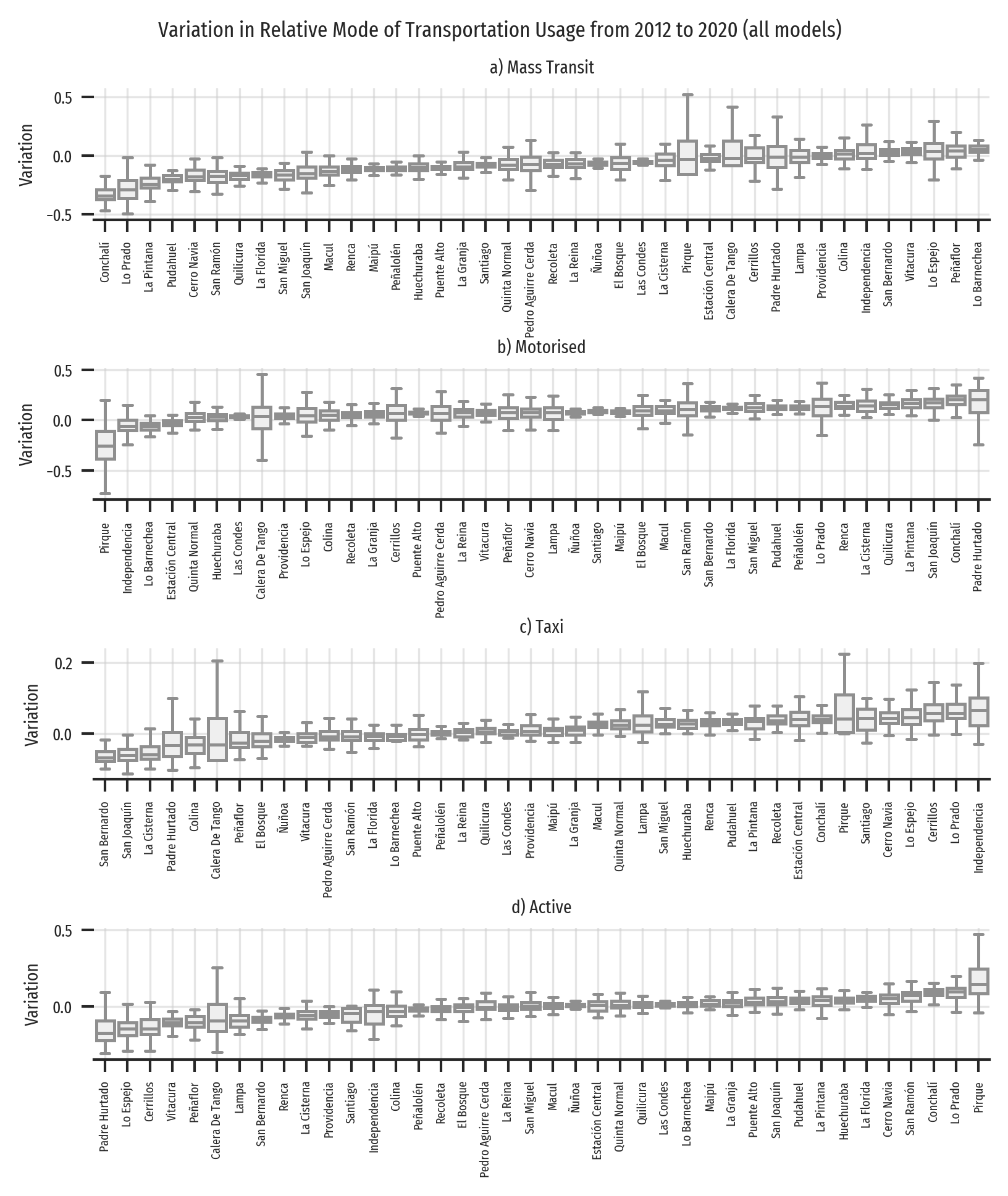}
\caption{Each boxplot shows how the relative share of the mode split per
municipality varies with respect to 100 adjusted models. Each model has
a different random initialization.}\label{fig:model_variability}
}
\end{figure}

In total, we adjusted 100 different models with different seeds for the
random initialization. Here, we report the reconstruction of the
\(R01'\) matrix in all these models (see Figure
\ref{fig:model_variability}). We observe that the relative mode share
per municipality exhibits variability, however, the municipalities with
larger variability are outside of the main urban area. This could be
expected, as peripheral municipalities have less data available. In
general, the behaviour is arguably stable: the greatest standard
deviation in share change per mode of transportation is 0.12
(\emph{mass-transit}), whereas the smallest is 0.05 (\emph{taxi}).

\begin{figure}
\hypertarget{fig:model_error}{%
\centering
\includegraphics{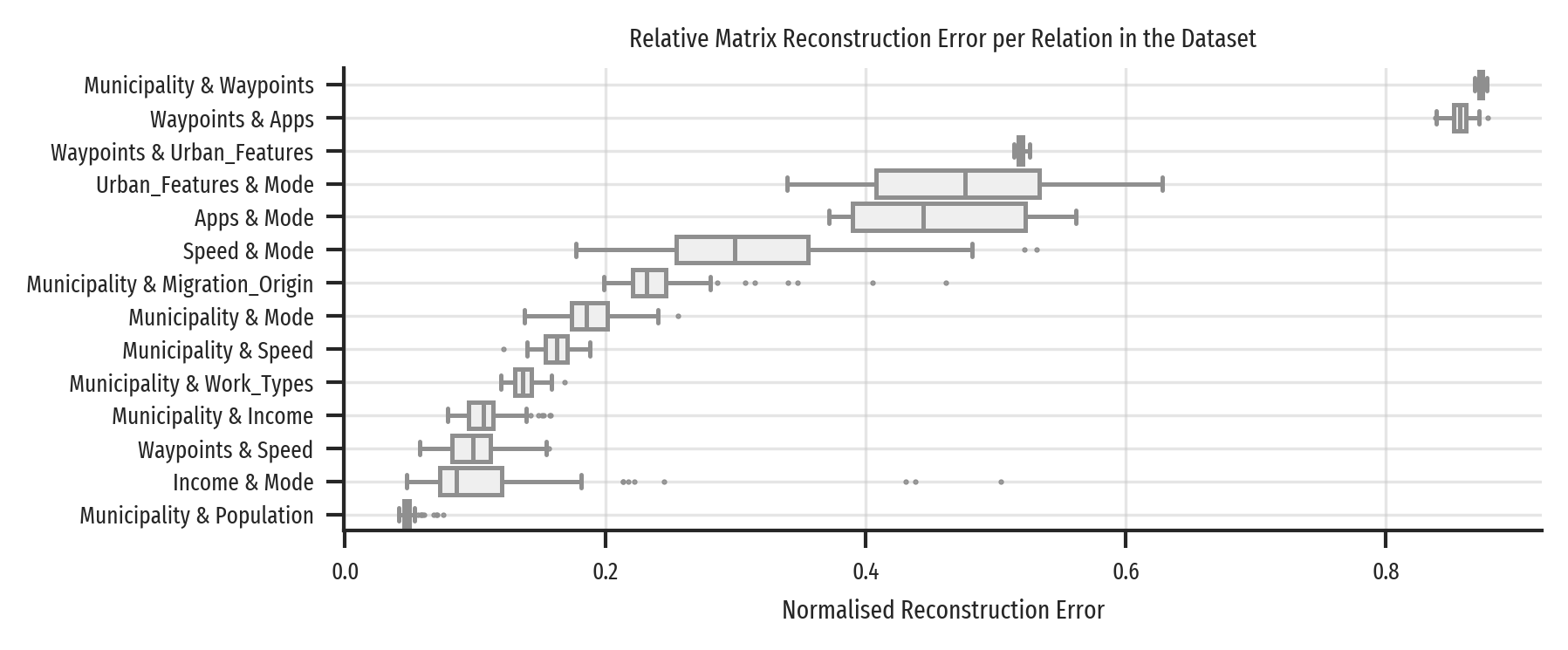}
\caption{Boxplot of analysis of model error. Each glyph represents the
normalised reconstruction error for a relation in the dataset. A box
comprises the range between the first and third quartile of values, as
well as the median of the distribution. Whisker lines express the spread
of the data, and dots represent outlier model instances. In total, we
ran 100 model instances.}\label{fig:model_error}
}
\end{figure}

Regarding reconstruction error per matrix, there are several magnitudes
depending on the relation being analysed (see Figure
\ref{fig:model_error}). First, the matrices related to population
statistics had a very small error (median of 0.05 in all fitted models),
whereas other matrices presented slightly higher values (such as the
median of 0.11 in the municipality and income relation). This implies
that the latent representations of the official data from the city are
accurate. Moderate errors come from matrices such as the relation
between application usage and mode of transportation (median of 0.44).
This value was expected to be high, as few applications were known to be
associated with modes of transportation, and the remaining applications
had a uniform imputed value of 0.25 (so they are equally associated with
each mode of transportation). Larger errors come from the mobile phone
data relations (median of 0.87 for the municipality and waypoints
relation), which are larger and sparser than the other relations. As
such, this was expected, as these relations exhibit detailed behaviour
from the population at waypoint resolution, whereas the other matrices,
which are used to reconstruct the data, contain data at a coarser
resolution. We expect the model to find the most important patterns
connected to the other matrices, as in previous work using matrix
factorization of mobile phone data
(\protect\hyperlink{ref-graells2018inferring}{Graells-Garrido et al.,
2018b}). To select the best model among the 100 instances, we chose the
one with the lowest global reconstruction error, defined as the
geometric mean of individual reconstruction errors. We observe that the
municipality and mode of transportation relation (\(R01'\)) exhibited a
median change of 0.2, which indicates a moderate change in
transportation patterns. Recall that this change is interpreted as the
change in the modal split rather than error. We discuss these changes
for the selected model in the next section. Since we focus the remainder
of this paper in the updated mode split by this model and its
validation, we refer the reader to Supplementary Material Section A4 for
a detailed view of all latent matrices for concepts and relationships.

\hypertarget{updated-mode-split}{%
\subsection{Updated Mode Split}\label{updated-mode-split}}

Next, we used the best model to reconstruct an updated mode split for
the city. Compared with the initial mode split from 2012, we observe the
following changes from the best factorisation model: a decrease in
\emph{mass-transit} usage from 39.44\% to 31.40\%, an increase in
\emph{motorised} transport from 40.64\% to 50.36\%, a decrease in
\emph{active} trips from 16.15\% to 13.47\%; and an increase in
\emph{taxi} trips 3.76\% to 4.76\%. These changes are similar to those
observed between the Santiago Travel Survey 2012 and 2001:
\emph{motorised} transport increased by 5.1\% and \emph{mass-transit}
was reduced by 6.4\% (\protect\hyperlink{ref-munoz2015encuesta}{Muñoz et
al., 2015}).

\begin{figure}
\hypertarget{fig:change_map}{%
\centering
\includegraphics{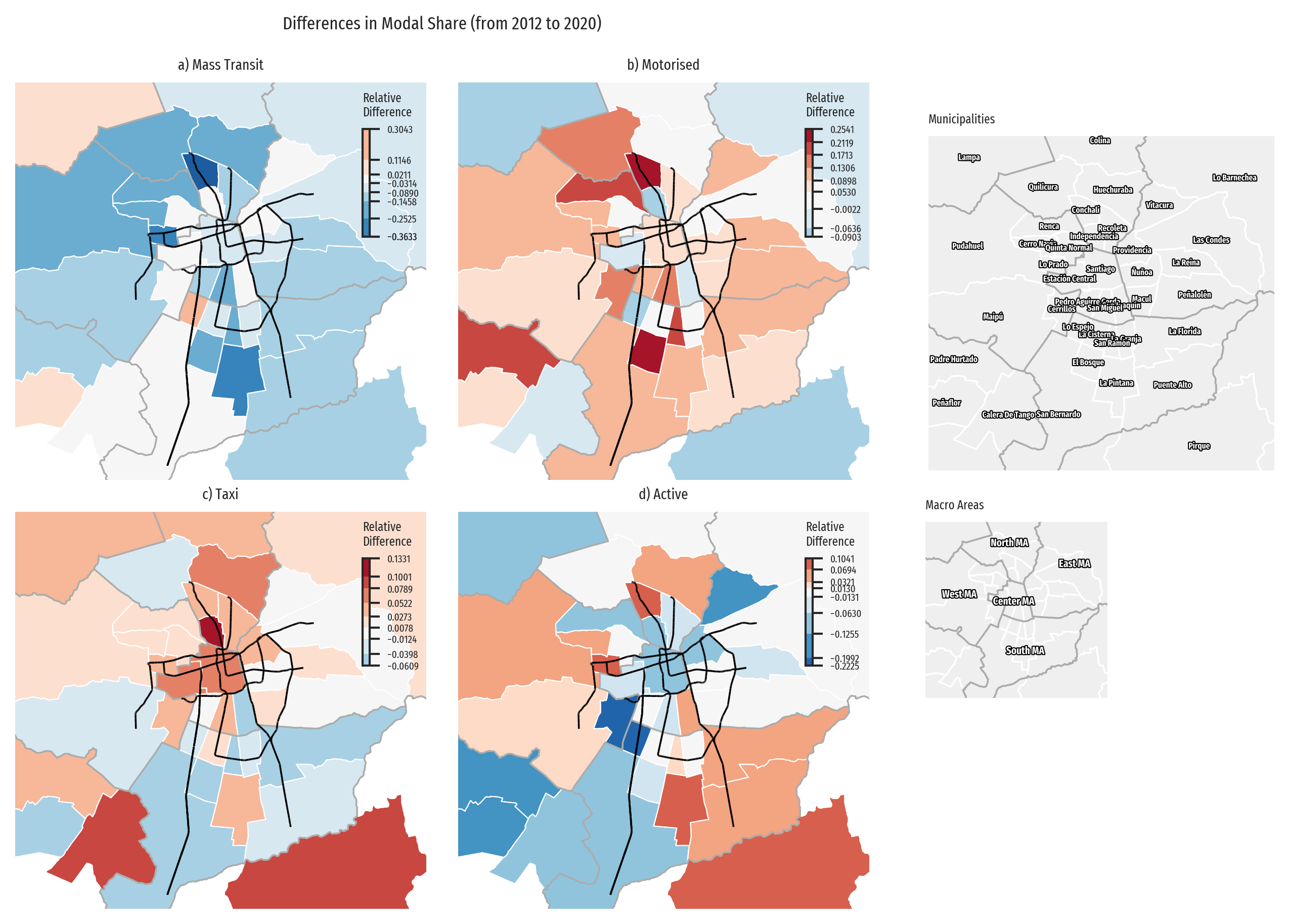}
\caption{Geographical distribution of modal split differences per
municipality.}\label{fig:change_map}
}
\end{figure}

The differences in the mode split reconstruction exhibit a geographical
dependence on the distribution (see Figure \ref{fig:change_map}). Most
relevant changes happen in municipalities crossed or surrounded by metro
or train lines. Only three municipalities exhibit an increase in
\emph{mass-transit} share: Lo Espejo, Lampa, and Peñaflor; of them, only
Lo Espejo is in the urban area. The increase in mass-transit share in Lo
Espejo is expected since Pineda and Lira
(\protect\hyperlink{ref-pineda2019travel}{2019}) found that after the
implementation of the metro line 6 in 2017, trips with origin in Lo
Espejo reduced their travel time by around 8 minutes. Additionally, this
municipality was benefited for the first time with a train station
integrated into the public transport system in 2017, whereas Lampa and
Peñaflor may have been benefited by their arguably close distance to the
terminal stations of the new train and metro services put into operation
in 2017 and 2019, respectively. Other municipalities exhibit a small
variation in \emph{mass-transit}, including Cerrillos, San Joaquín,
Estación Central, Lo Prado, Independencia, and Providencia. These
municipalities have access to new metro lines since 2017
(\protect\hyperlink{ref-pineda2019travel}{Pineda and Lira, 2019}). Of
particular interest is the \emph{mass-transit} increase in Vitacura, one
of the wealthiest municipalities in the country, with a strong share of
\emph{motorised} transport in the last travel survey. We hypothesise
that, due to the increase of remote work after the social outburst in
2019, the number of \emph{motorised} trips decreased, thus, increasing
the relative share of \emph{mass-transit}. We included a relationship
between municipalities and occupations to account for this phenomena in
our dataset. However, as there is not an official dataset regarding the
share of remote work for 2022, the test of this hypothesis is left for
future work. Conversely, Conchalí, in the northern area of the urban
radius, which is also crossed by a new metro line, exhibited a high
decrease in \emph{mass-transit} share; in turn, all other three modes
increased considerably. Metro operations in Conchalí started in 2019;
then, we hypothesise that the social upheaval from that year delayed the
expected modal shift to public transport.All other municipalities
decreased their \emph{mass-transit} share. Most of them were already
well-connected to public transport. This difference could be explained
by the poor quality of service of the public transport system, which is
globally perceived as overcrowded and slow
(\protect\hyperlink{ref-tirachini2017estimation}{Tirachini et al.,
2017}).

\emph{Motorised}, \emph{taxi} and \emph{active} trips exhibit an
increase through municipalities in the entire area of study. Arguably,
the increase in \emph{motorised} is the most expected relationship, as
cars have increased their sales in the last years (as evidenced by the
increase in car permits), and as these vehicles are likely to replace
public transportation for their owners
(\protect\hyperlink{ref-beirao2007understanding}{Beirão and Cabral,
2007}). A similar expectation holds with respect to \emph{taxi},
although there are specific cases worthy of discussion. The literature
states that ride-hailing applications tend to replace public transport
and traditional taxis in Santiago
(\protect\hyperlink{ref-tirachini2019ride}{Tirachini and Río, 2019}),
and \emph{taxi} exhibits a geographical pattern, as most of the
municipalities with increased \emph{taxi} share belong to the Center,
North and West macro areas of the city. These macro areas had a stronger
metro network than the others, thus, we may hypothesise that
ride-hailing has a tendency to replace metro trips in Santiago,
including municipalities with access to new metro lines. This result may
be controversial, as it is expected that the number of trips in public
transportation would increase in those municipalities served by the new
metro line since it generated a decrease of 14\% in travel time for its
users (\protect\hyperlink{ref-pineda2019travel}{Pineda and Lira, 2019});
thus, this opens a line for future work regarding the effect of the
social outburst in travel patterns in Santiago. Finally, one interesting
pattern for us is how the municipalities where \emph{active} transport
increased are primarily residential in the West and South macro areas.
Further work is needed to understand these changes, as this is an aspect
of transportation affected by the COVID-19 pandemic and its
corresponding municipal mobility restrictions
(\protect\hyperlink{ref-pappalardo2023dataset}{Pappalardo et al.,
2023}).

\hypertarget{model-validation-with-external-sources}{%
\subsection{Model Validation with External
Sources}\label{model-validation-with-external-sources}}

We made the first comparison between our results and ADATRAP at the
macro area level (see Table 3). We grouped the total averaged trips from
each macro area. We observe that there is a systematic difference, where
the proposed model always predicted less trips than ADATRAP. This could
be related to our initial modulation of \emph{bip!} smart card registers
from the National Institute of Statistics, which could have been
stronger, and thus, the model would have reported more
\emph{mass-transit} trips. Then, to quantify how similar the prediction
of our model was to the actual measured trips at municipality level, we
used the Pearson correlation coefficient \(r\), defined as follows:
\begin{equation}
r = \frac{\text{cov}(X,Y)}{\sigma_{X} \sigma_{Y}}.
\end{equation} This coefficient measures the statistical association
between two vectors with the municipal averaged trip counts (\(X\) and
\(Y\)). The \(r\) coefficient lies in \([-1, 1]\), where a value of -1
indicates negative correlation, a value of 1 indicates positive
correlation, and 0 indicates total independence. The results obtained
from the comparison yielded a value of \(r =\) 0.88 (\(p <\) 0.001) for
the proposed model, suggesting a strong positive correlation between the
predicted trips and the actual measured trips. The initial guess of the
updated mode split also had a correlation of \(r =\) 0.84 (\(p <\)
0.001), indicating that the initial estimation of \emph{mass-transit}
share was fairly accurate. The proposed model captures behaviour
observed in official data sources and improves the initial guess by
integrating digital traces and other official data.

\begin{longtable}[]{@{}
  >{\raggedright\arraybackslash}p{(\columnwidth - 4\tabcolsep) * \real{0.2973}}
  >{\raggedright\arraybackslash}p{(\columnwidth - 4\tabcolsep) * \real{0.3649}}
  >{\raggedright\arraybackslash}p{(\columnwidth - 4\tabcolsep) * \real{0.3108}}@{}}
\caption{Comparison of average daily commuting trips between our
proposed model and those reported by the ADATRAP system.}\tabularnewline
\toprule()
\begin{minipage}[b]{\linewidth}\raggedright
Macro area
\end{minipage} & \begin{minipage}[b]{\linewidth}\raggedright
Proposed model
\end{minipage} & \begin{minipage}[b]{\linewidth}\raggedright
ADATRAP
\end{minipage} \\
\midrule()
\endfirsthead
\toprule()
\begin{minipage}[b]{\linewidth}\raggedright
Macro area
\end{minipage} & \begin{minipage}[b]{\linewidth}\raggedright
Proposed model
\end{minipage} & \begin{minipage}[b]{\linewidth}\raggedright
ADATRAP
\end{minipage} \\
\midrule()
\endhead
Center & 106,969 & 182,790 \\
North & 80,261 & 107,999 \\
East & 129,314 & 184,268 \\
West & 137,696 & 212,244 \\
South & 190,424 & 244,946 \\
Total & 644,664 & 932,247 \\
\bottomrule()
\end{longtable}

\begin{figure}
\hypertarget{fig:validation}{%
\centering
\includegraphics{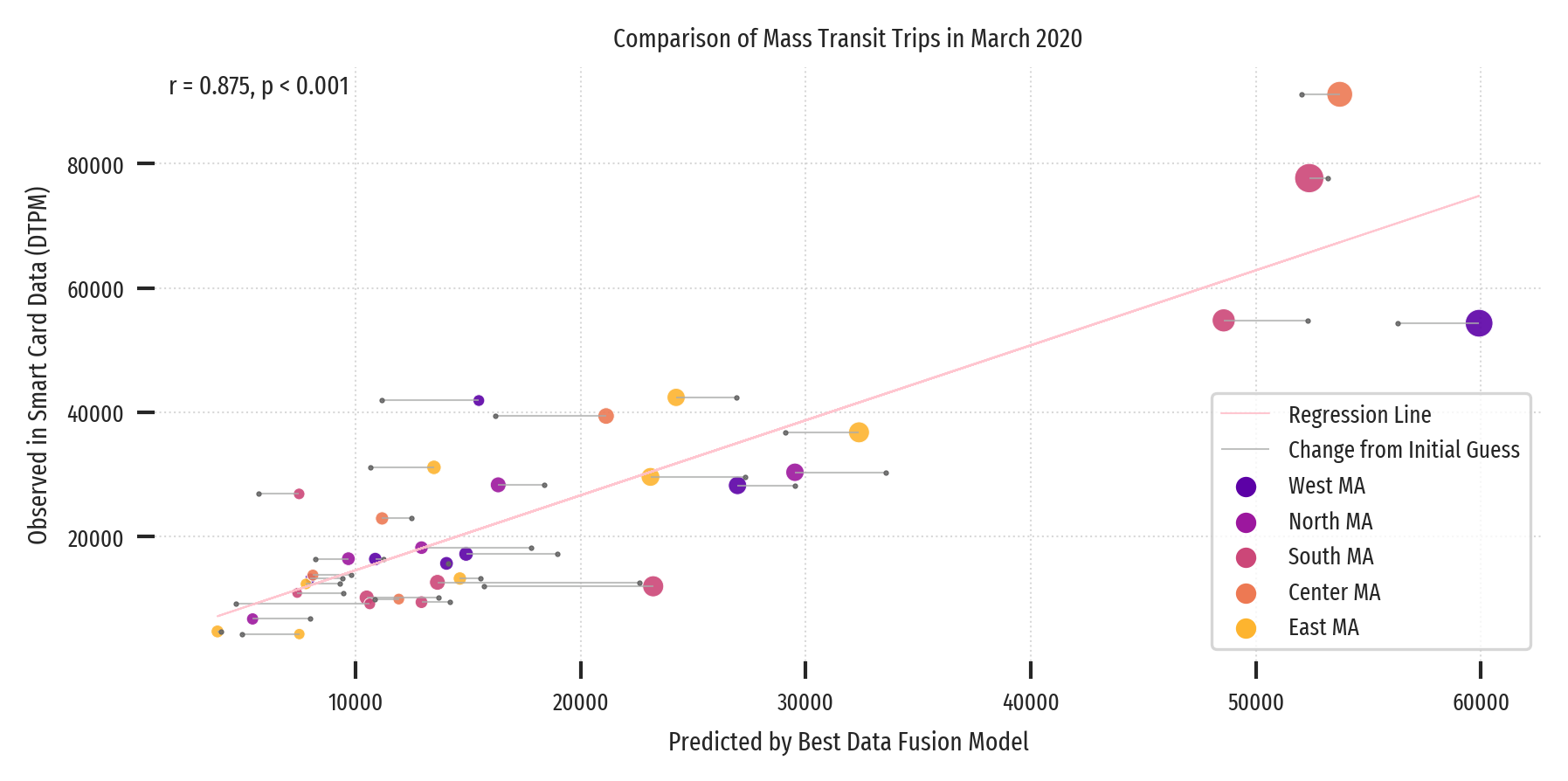}
\caption{The bubble plot of correlation shows the relationship between
the mass-transit trip counts predicted by the proposed model and the
actual counts observed in smart card data for the same period of study.
Each bubble in the plot represents a municipality and its size is
proportional to the municipality's population in 2020. In addition, each
bubble is connected to a small dot that represents the initial
\emph{mass-transit} trip count used by the model.}\label{fig:validation}
}
\end{figure}

\begin{figure}
\hypertarget{fig:errors}{%
\centering
\includegraphics{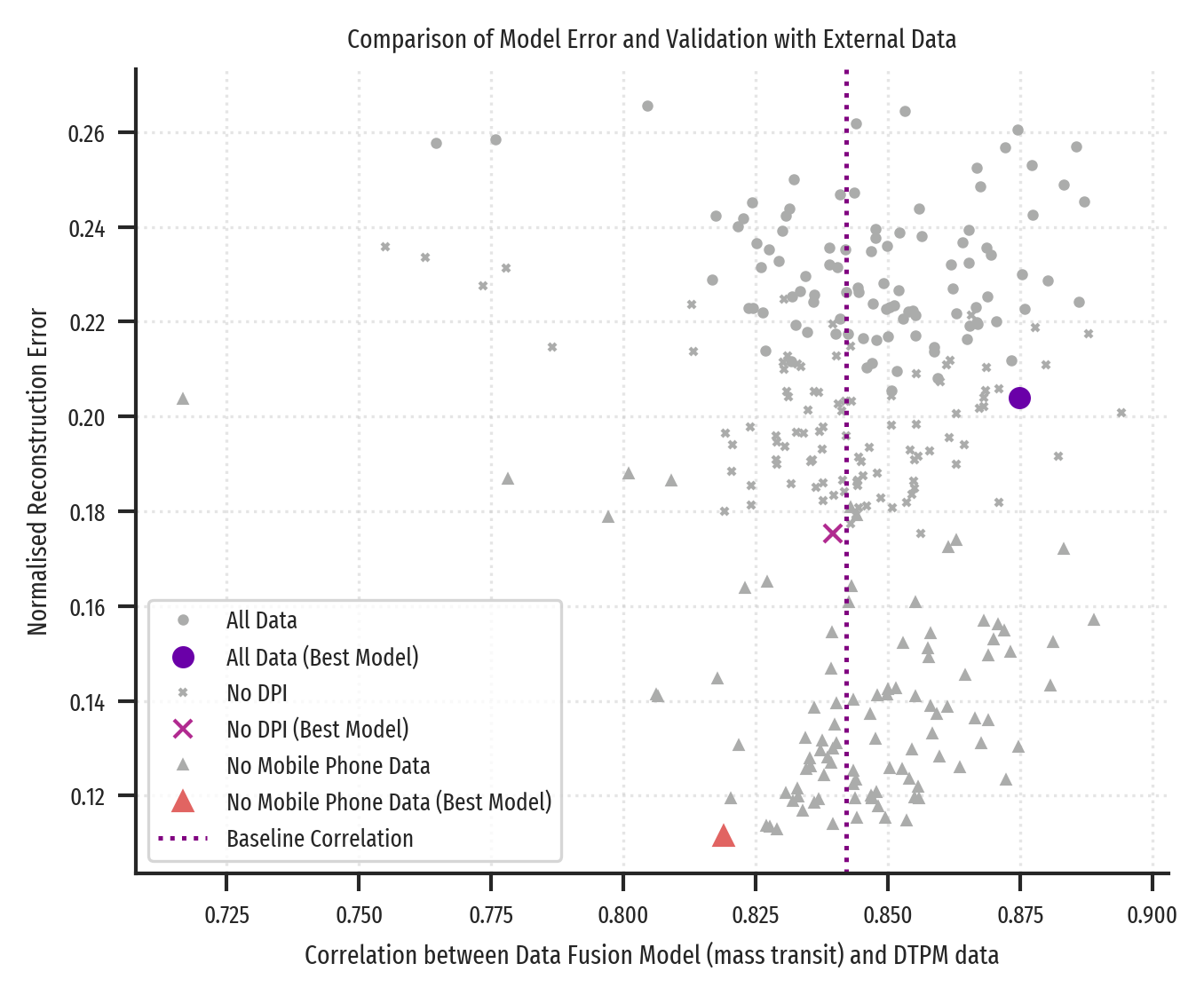}
\caption{Relationship between model error and correlation between model
outcomes for mass transit and validation data. We tested three data
configurations: the proposed one that fuses all available data (official
data, OpenStreetMap and mobile phone data from XDR and DPI), one with
the same data with exception of DPI (application usage), and one using
only official data and OSM. Each marker represents a model within each
group, with the best model of each configuration highlighted with a
greater size and with colour.}\label{fig:errors}
}
\end{figure}

On the other hand, since we integrated several official data sources,
where some of them are not directly related to transportation, we tested
two additional data configurations: one without DPI data, and without
any mobile phone data (XDR and DPI), with all the other sources
remaining active in each configuration. Note that the process to adjust
a model in each configuration is always the same, the only difference is
the amount of available relation matrices. For each configuration, we
fitted 100 model instances, and then we estimated the Pearson
correlation coefficient \(r\) between each instance and the observed
data by ADATRAP (see Figure \ref{fig:errors}). We observed that the
global error is always greater when using more data. This can be
explained due to the size and density of the relationships that involve
waypoints (mobile phone towers), which are one order of magnitude
greater than the rest of the data. We also observe that our criteria to
select the best model does not always pick the model with the best
correlation with ADATRAP data. In the configuration with all data
available, the best model is one of the most correlated with observed
data. However, in the other configurations, this is not guaranteed;
indeed, in the configuration without mobile phone data, the best model
has one of the worst correlations with observed data. Additionally, the
best models for the test configurations present a worse result than the
baseline correlation obtained with the initial guess of the mode split.
Thus, even though our best model criteria does not find the actual best
model according to ground truth data, it does pick one that is
considerably better than the baseline. This suggests that future work
should identify an ad-hoc strategy for model selection.

Finally, one of the contributions from this work is the usage of DPI
(application usage) from mobile phones. Then, it would be important to
assess the effect of incorporating this source into the model. The
existence of a potentially better or comparable result using less data
requires additional analysis to justify the usage of several types of
mobile phone data. At this point, we already know that one quality is
its determinism, that is, being able to select the best model using an
established criteria based on global error. However, we also need to
identify how different the results are. To do so, we estimated the
Pearson correlation coefficients for each mode split between our
proposal and the other two data configurations (see Table 4). We observe
that the mode splits are similar between configurations with mobile
phone data, with high correlation coefficients (all greater or equal
than 0.9, with statistical significance). We also observe that the only
mode of transportation that presents a divergence from the obtained mode
splits is \emph{taxi} when working without mobile phone data (0.49,
\(p =\) 0.001). This has a two-fold implication. On the one hand, it
becomes clear that the contribution of the mobile phone data data is the
ability to pinpoint the new \emph{taxi} share, arguably through the
signal of ride-hailing application usage. For all other modes, the
official data used may have latent factors that enable the model to
identify part of their changes, however, such analysis should be
approached in future work. On the other hand, although this result
suggests that without using DPI (app usage), or even XDR (trajectories),
we found that it is possible to have a good up-to-date estimation of the
modal partition. We highlight that the two types of mobile phone data
are necessary, as only when using them we have a replicable criteria to
find a model with good results in comparison to the baseline comparison.

\begin{longtable}[]{@{}
  >{\raggedright\arraybackslash}p{(\columnwidth - 4\tabcolsep) * \real{0.2838}}
  >{\raggedright\arraybackslash}p{(\columnwidth - 4\tabcolsep) * \real{0.2838}}
  >{\raggedright\arraybackslash}p{(\columnwidth - 4\tabcolsep) * \real{0.4054}}@{}}
\caption{Pearson correlation coefficients between the predicted mode
split in the proposed data configuration and two configurations with
less data but similar validation results. All p-values are
Bonferroni-corrected.}\tabularnewline
\toprule()
\begin{minipage}[b]{\linewidth}\raggedright
\end{minipage} & \begin{minipage}[b]{\linewidth}\raggedright
No DPI
\end{minipage} & \begin{minipage}[b]{\linewidth}\raggedright
No Mobile Phone Data
\end{minipage} \\
\midrule()
\endfirsthead
\toprule()
\begin{minipage}[b]{\linewidth}\raggedright
\end{minipage} & \begin{minipage}[b]{\linewidth}\raggedright
No DPI
\end{minipage} & \begin{minipage}[b]{\linewidth}\raggedright
No Mobile Phone Data
\end{minipage} \\
\midrule()
\endhead
\emph{Mass Transit} & 0.98 (p \textless{} 0.001) 0 & .97 (p \textless{}
0.001) \\
\emph{Motorised} & 0.98 (p \textless{} 0.001) 0 & .97 (p \textless{}
0.001) \\
\emph{Taxi} & 0.90 (p = 0.4) & 0.49 (p = 0.001) \\
\emph{Active} & 0.95 (p \textless{} 0.001) 0 & .93 (p \textless{}
0.001) \\
\bottomrule()
\end{longtable}

\hypertarget{discussion}{%
\section{Discussion}\label{discussion}}

In practice, the Data Fusion approach requires a limited amount of data
(a weekday of mobile phone data, as in the case study) and supports
incorporating domain knowledge or any other data source that can be
expressed as a matrix. This is a strength of the proposed method, as the
demand for frameworks to fuse data from a multidisciplinary perspective
is increasing (\protect\hyperlink{ref-meta2021physiology}{Meta et al.,
2022}), and incorporating domain knowledge is crucial to the adoption of
data-driven initiatives for transportation planning
(\protect\hyperlink{ref-graells2020adoption}{Graells-Garrido et al.,
2020}). Another strength is that the model works with aggregated data.
Although we used individual-level mobile phone data, the actual input
for the model was mobility behaviour aggregated at administrative levels
(for XDR trajectory data), and at cell phone tower levels (for DPI
application usage data). The usage of aggregated data arguably
diminishes potential issues regarding privacy, and facilitates data
transfer between entities, due to its reduced size and its inherent
anonymisation. Furthermore, the good correlations observed with models
using different subsets of the available data indicate that the data
fusion approach has potential even when using official data and OSM
only, which also opens a future line of work to identify which already
available official sources can aid in the determination of an updated
mode split.

The main theoretical implication of this work is that data-driven
transportation planning projects can still provide relevant insights
even in the presence of biassed data. Data bias is commonly considered
an artefact affecting digital traces, such as the mobile phone data
integrated here. Such bias comes from access to technology due to
income, cultural, social, or geographical barriers. Note that the model
implicitly accounts for this because the integration of official
socio-economic data affects how the model computes the representation of
each concept in the dataset: the mobile phone data is used as one
additional data source, not the main one in estimating the mode split.
We find this relevant because assessing bias for mobile phone data is
challenging, due to the data being anonymised. Indeed, despite the
presence of bias in the digital traces, the correlation obtained by our
model is high and statistically significant.

\hypertarget{limitations}{%
\subsection{Limitations}\label{limitations}}

We identify three main limitations in this work. First, our validation
only concerns the public transport share, although the observed
differences in other modes of transportation are coherent with
expectations from the literature and the urban context. However, to
assess all the model outcomes, a comparison with official data that
includes all the other modes of transportation must be made. To the
extent of our knowledge, the data necessary to make such a comparison
does not exist at the moment of writing this paper.

Second, our method integrates official data regarding socio-economic
characteristics, occupations, migration, and the local population.
Arguably these sources implicitly control for data bias, yet the extent
of such control is not quantified. This aspect is relevant mainly for
policymakers and urban planners, as they would base their work on the
results reported by data-driven models such as the one presented here.
We also observe that the temporal granularity of the data we analysed is
different, however, we argue that the phenomena under study (mode split)
is depicted with the finest detail possible, as we are using mobile
phone data, whereas official sources may represent yearly averages of
phenomena that is well captured by the official sources as, indeed, the
authority and other institutions make use of that data every year.

Third, our results are bound by the mobile phone network event frequency
and the geographical tower distribution. Connected phones could be
redirected between towers if this is needed to handle network requests,
which could result in false positives in trip detection. However, XDR is
an aggregated dataset that consolidates multiple connections into a
single record, with an interval of 15 minutes in particular in our
context. Although we cannot control by this problem, as the XDR data is
already aggregated, we expect its effect to be negligible in our setting
due to the consolidation of the data, the fact that in our study period
the expected behaviour of people is to travel to work or study
locations, and that towers are distributed in the city to ensure quality
of service (as reported by the mobile phone operator). As a result of
this limitation, a direct application of this work into other countries
or datasets may need a sensitivity analysis to determine if mobility
patterns inferred from XDR (or similar) have enough resolution to be
included in the Data Fusion model.

\hypertarget{future-work}{%
\subsection{Future Work}\label{future-work}}

There is still a space to be explored regarding the variability of
outcomes given different initial conditions, as well as the solution
space of the proposed method. On the one hand, the initial solution used
as base for the Data Fusion process provided a solid starting point,
however, the method is bound by these initial conditions; a different
strategy to build the initial solution may provide other results. Yet,
our validation proves that this was, indeed, a well defined initial
condition. On the other hand, our process to select a model did not
perform any regularisation or reduction of the solution space, as it was
based only on the random initialization of the latent matrices. One
potential way to subvert this limitation is to include an additional
operation in the optimization problem: restriction of must-link and
cannot-link values of each concept. This operation is supported by the
original model (\protect\hyperlink{ref-vzitnik2014data}{Žitnik and
Zupan, 2014}). However, the specification of these restrictions needs
additional domain knowledge. We leave the exploration of that solution
to future work.

Our work focuses on commuting times, the most recurring activity in a
city, however, there are many other activities and time periods that
should be accounted for in urban planning. We expect that the model
could be made temporally-aware by working with tensors in all
relationships that may have a temporal dimension, such as the aggregated
trajectories or the application usage per municipality
(\protect\hyperlink{ref-chen2019missing}{Chen et al., 2019}).

\hypertarget{conclusions}{%
\section{Conclusions}\label{conclusions}}

In this study, we developed a novel methodology to address the data
challenges in evaluating and informing transport planning interventions
in cities. By utilising mobile phone data and data fusion techniques, we
accurately estimated the mode of transportation usage in Santiago, a
major Latin American city.

Our findings reveal significant shifts in transportation modes over
time. We observed a decrease in mass-transit usage, except in areas with
new metro/rail lines, indicating the impact of these infrastructure
improvements. Motorised transport has increased citywide, highlighting
the ongoing challenges in promoting sustainability. Additionally, taxi
usage, including traditional taxis and ride-hailing applications, has
risen, while the share of active transportation, such as walking and
cycling, has declined. We validated these results using official data
from smart card transactions, demonstrating the reliability and accuracy
of our methodology. The consistency of our findings with domain
knowledge and historical trends further supports the robustness of our
approach.

Looking ahead, as cities face escalating challenges in achieving
sustainable transportation goals, cost-effective tools providing finer
data resolutions will be essential. Collaboration among scientists,
private and public institutions, and the continued development of
innovative methods, like ours, will play a pivotal role in addressing
these challenges effectively.

In conclusion, our research advances the understanding of travel
behaviour patterns and underscores the need for data-driven approaches
in sustainable transport planning. By harnessing mobile phone data and
data fusion techniques, cities can make informed decisions to promote
public transit, reduce reliance on motorised vehicles, and foster a more
sustainable and resilient urban future.

\hypertarget{references}{%
\section*{References}\label{references}}
\addcontentsline{toc}{section}{References}

\hypertarget{refs}{}
\begin{CSLReferences}{1}{0}
\leavevmode\vadjust pre{\hypertarget{ref-acar2020potential}{}}%
Acar, C., Dincer, I., 2020. The potential role of hydrogen as a
sustainable transportation fuel to combat global warming. International
Journal of Hydrogen Energy 45, 3396--3406.

\leavevmode\vadjust pre{\hypertarget{ref-policymakersbook2012}{}}%
Allen, S.K., Barros, V., Burton, I., Campbell-Lendrum, D., Cardona,
O.-D., Cutter, S.L., Dube, O.P., Ebi, K.L., Field, C.B., Handmer, J.W.,
al., et, 2012. Summary for policymakers, in: Field, C.B., Barros, V.,
Stocker, T.F., Dahe, Q. (Eds.), Managing the Risks of Extreme Events and
Disasters to Advance Climate Change Adaptation: Special Report of the
Intergovernmental Panel on Climate Change. Cambridge University Press,
pp. 3--22. \url{https://doi.org/10.1017/CBO9781139177245.003}

\leavevmode\vadjust pre{\hypertarget{ref-antoniou2011synthesis}{}}%
Antoniou, C., Balakrishna, R., Koutsopoulos, H.N., 2011. A synthesis of
emerging data collection technologies and their impact on traffic
management applications. European Transport Research Review 3, 139--148.

\leavevmode\vadjust pre{\hypertarget{ref-asgari2016inferring}{}}%
Asgari, F., 2016. Inferring user multimodal trajectories from cellular
network metadata in metropolitan areas (PhD thesis). Institut National
des T{é}l{é}communications.

\leavevmode\vadjust pre{\hypertarget{ref-bachir2019mode}{}}%
Bachir, D., Khodabandelou, G., Gauthier, V., El Yacoubi, M., Puchinger,
J., 2019. Inferring dynamic origin-destination flows by transport mode
using mobile phone data. Transportation Research Part C: Emerging
Technologies 101, 254--275.
\url{https://doi.org/10.1016/j.trc.2019.02.013}

\leavevmode\vadjust pre{\hypertarget{ref-bantis2017you}{}}%
Bantis, T., Haworth, J., 2017. Who you are is how you travel: A
framework for transportation mode detection using individual and
environmental characteristics. Transportation Research Part C: Emerging
Technologies 80, 286--309.

\leavevmode\vadjust pre{\hypertarget{ref-beirao2007understanding}{}}%
Beirão, G., Cabral, J.S., 2007. Understanding attitudes towards public
transport and private car: A qualitative study. Transport policy 14,
478--489.

\leavevmode\vadjust pre{\hypertarget{ref-breyer2022}{}}%
Breyer, N., Rydergren, C., Gundlegård, D., 2022. Semi-supervised mode
classification of inter-city trips from cellular network data. Journal
of Big Data Analytics in Transportation 1--17.

\leavevmode\vadjust pre{\hypertarget{ref-browning2006effects}{}}%
Browning, R.C., Baker, E.A., Herron, J.A., Kram, R., 2006. Effects of
obesity and sex on the energetic cost and preferred speed of walking.
Journal of applied physiology 100, 390--398.

\leavevmode\vadjust pre{\hypertarget{ref-brueckner2005transport}{}}%
Brueckner, J.K., 2005. Transport subsidies, system choice, and urban
sprawl. Regional Science and Urban Economics 35, 715--733.

\leavevmode\vadjust pre{\hypertarget{ref-calabrese2011}{}}%
Calabrese, F., Colonna, M., Lovisolo, P., Parata, D., Ratti, C., 2011.
Real-time urban monitoring using cell phones: A case study in rome. IEEE
Transactions on Intelligent Transportation Systems 12, 141--151.
\url{https://doi.org/10.1109/TITS.2010.2074196}

\leavevmode\vadjust pre{\hypertarget{ref-chang2019modechoice}{}}%
Chang, X., Wu, J., Liu, H., Yan, X., Sun, H., Qu, Y., 2019. Travel mode
choice: A data fusion model using machine learning methods and evidence
from travel diary survey data. Transportmetrica A: Transport Science 15,
1587--1612. \url{https://doi.org/10.1080/23249935.2019.1620380}

\leavevmode\vadjust pre{\hypertarget{ref-chen2017mago}{}}%
Chen, K.-Y., Shah, R.C., Huang, J., Nachman, L., 2017. Mago: Mode of
transport inference using the hall-effect magnetic sensor and
accelerometer. Proceedings of the ACM on Interactive, Mobile, Wearable
and Ubiquitous Technologies 1, 1--23.

\leavevmode\vadjust pre{\hypertarget{ref-chen2019missing}{}}%
Chen, X., He, Z., Chen, Y., Lu, Y., Wang, J., 2019. Missing traffic data
imputation and pattern discovery with a bayesian augmented tensor
factorization model. Transportation Research Part C: Emerging
Technologies 104, 66--77.

\leavevmode\vadjust pre{\hypertarget{ref-colvile2001transport}{}}%
Colvile, R., Hutchinson, E.J., Mindell, J., Warren, R., 2001. The
transport sector as a source of air pollution. Atmospheric environment
35, 1537--1565.

\leavevmode\vadjust pre{\hypertarget{ref-vcopar2019fast}{}}%
Čopar, A., Zupan, B., Zitnik, M., 2019. Fast optimization of
non-negative matrix tri-factorization. PloS one 14, e0217994.

\leavevmode\vadjust pre{\hypertarget{ref-dabiri2018inferring}{}}%
Dabiri, S., Heaslip, K., 2018. Inferring transportation modes from GPS
trajectories using a convolutional neural network. Transportation
research part C: emerging technologies 86, 360--371.

\leavevmode\vadjust pre{\hypertarget{ref-ding2006tnmf}{}}%
Ding, C., Li, T., Peng, W., Park, H., 2006. Orthogonal nonnegative
matrix t-factorizations for clustering, in: Proceedings of the 12th ACM
SIGKDD International Conference on Knowledge Discovery and Data Mining,
KDD '06. Association for Computing Machinery, New York, NY, USA, pp.
126--135. \url{https://doi.org/10.1145/1150402.1150420}

\leavevmode\vadjust pre{\hypertarget{ref-doyle2011utilising}{}}%
Doyle, J., Hung, P., Kelly, D., McLoone, S.F., Farrell, R., 2011.
Utilising mobile phone billing records for travel mode discovery.

\leavevmode\vadjust pre{\hypertarget{ref-el2011data}{}}%
El Faouzi, N.-E., Leung, H., Kurian, A., 2011. Data fusion in
intelligent transportation systems: Progress and challenges--a survey.
Information Fusion 12, 4--10.

\leavevmode\vadjust pre{\hypertarget{ref-garcia2016big}{}}%
García, P., Herranz, R., Javier, J., 2016. Big data analytics for a
passenger-centric air traffic management system. 6th SESAR Innovation
Days.

\leavevmode\vadjust pre{\hypertarget{ref-xu2019mining}{}}%
González, M.C., 2019. Mining urban lifestyles: Urban computing, human
behavior and recommender systems. Computing.
\url{https://doi.org/10.1049/PBPC035G_ch5}

\leavevmode\vadjust pre{\hypertarget{ref-graells2018and}{}}%
Graells-Garrido, E., Caro, D., Miranda, O., Schifanella, R., Peredo,
O.F., 2018a. The {WWW} (and an {H}) of mobile application usage in the
city: The what, where, when, and how, in: Companion Proceedings of the
the Web Conference 2018. pp. 1221--1229.

\leavevmode\vadjust pre{\hypertarget{ref-graells2018inferring}{}}%
Graells-Garrido, E., Caro, D., Parra, D., 2018b. Inferring modes of
transportation using mobile phone data. EPJ Data Science 7, 49.

\leavevmode\vadjust pre{\hypertarget{ref-graells2017effect}{}}%
Graells-Garrido, E., Ferres, L., Caro, D., Bravo, L., 2017. The effect
of {P}ok{é}mon {G}o on the pulse of the city: A natural experiment. EPJ
Data Science 6, 1--19.

\leavevmode\vadjust pre{\hypertarget{ref-graells2020adoption}{}}%
Graells-Garrido, E., Peña-Araya, V., Bravo, L., 2020. Adoption-driven
data science for transportation planning: Methodology, case study, and
lessons learned. Sustainability 12.
\url{https://doi.org/10.3390/su12156001}

\leavevmode\vadjust pre{\hypertarget{ref-graells2016sensing}{}}%
Graells-Garrido, E., Peredo, O., García, J., 2016. Sensing urban
patterns with antenna mappings: The case of {S}antiago, {C}hile. Sensors
16, 1098.

\leavevmode\vadjust pre{\hypertarget{ref-graells2022measuring}{}}%
Graells-Garrido, E., Schifanella, R., Opitz, D., Rowe, F., 2022.
Measuring the local complementarity of population, amenities and digital
activities to identify and understand urban areas of interest.
Environment and Planning B: Urban Analytics and City Science
23998083221117830.

\leavevmode\vadjust pre{\hypertarget{ref-grantz2020use}{}}%
Grantz, K.H., Meredith, H.R., Cummings, D.A., Metcalf, C.J.E., Grenfell,
B.T., Giles, J.R., Mehta, S., Solomon, S., Labrique, A., Kishore, N.,
others, 2020. The use of mobile phone data to inform analysis of
COVID-19 pandemic epidemiology. Nature communications 11, 4961.

\leavevmode\vadjust pre{\hypertarget{ref-green2021new}{}}%
Green, M., Pollock, F.D., Rowe, F., 2021. New forms of data and new
forms of opportunities to monitor and tackle a pandemic, in: COVID-19
and Similar Futures: Pandemic Geographies. Springer, pp. 423--429.

\leavevmode\vadjust pre{\hypertarget{ref-Gschwender2016vi}{}}%
Gschwender, A., Munizaga, M., Simonetti, C., 2016.
\href{http://www.sciencedirect.com/science/article/pii/S0739885915300998}{{Using
smart card and {GPS} data for policy and planning: The case of
{T}ransantiago}}. Competition and Ownership in Land Passenger Transport
(selected papers from the Thredbo 14 conference) 59 IS -, 242--249.

\leavevmode\vadjust pre{\hypertarget{ref-haklay2010good}{}}%
Haklay, M., 2010. How good is volunteered geographical information? A
comparative study of OpenStreetMap and ordnance survey datasets.
Environment and planning B: Planning and design 37, 682--703.

\leavevmode\vadjust pre{\hypertarget{ref-heinen2015same}{}}%
Heinen, E., Chatterjee, K., 2015. The same mode again? An exploration of
mode choice variability in great britain using the national travel
survey. Transportation Research Part A: Policy and Practice 78,
266--282.

\leavevmode\vadjust pre{\hypertarget{ref-holleczek2015traffic}{}}%
Holleczek, T., The Anh, D., Yin, S., Jin, Y., Antonatos, S., Goh, H.L.,
Low, S., Shi-Nash, A., 2015. Traffic measurement and route
recommendation system for mass rapid transit (mrt), in: Proceedings of
the 21th ACM SIGKDD International Conference on Knowledge Discovery and
Data Mining. pp. 1859--1868.

\leavevmode\vadjust pre{\hypertarget{ref-horn2017qztool}{}}%
Horn, C., Gursch, H., Kern, R., Cik, M., 2017. QZTool---automatically
generated origin-destination matrices from cell phone trajectories, in:
Advances in Human Aspects of Transportation: Proceedings of the AHFE
2016 International Conference on Human Factors in Transportation, July
27-31, 2016, Walt Disney World{\textregistered}, Florida, USA. Springer,
pp. 823--833.

\leavevmode\vadjust pre{\hypertarget{ref-horn2015deriving}{}}%
Horn, C., Kern, R., 2015. Deriving public transportation timetables with
large-scale cell phone data. Procedia Computer Science 52, 67--74.

\leavevmode\vadjust pre{\hypertarget{ref-huang2019transport}{}}%
Huang, H., Cheng, Y., Weibel, R., 2019. Transport mode detection based
on mobile phone network data: A systematic review. Transportation
Research Part C: Emerging Technologies 101, 297--312.

\leavevmode\vadjust pre{\hypertarget{ref-huang2018modeling}{}}%
Huang, Z., Ling, X., Wang, P., Zhang, F., Mao, Y., Lin, T., Wang, F.-Y.,
2018. Modeling real-time human mobility based on mobile phone and
transportation data fusion. Transportation research part C: emerging
technologies 96, 251--269.

\leavevmode\vadjust pre{\hypertarget{ref-hui2017investigating}{}}%
Hui, K.T.Y., Wang, C., Kim, A., Qiu, T.Z., 2017. Investigating the use
of anonymous cellular phone data to determine intercity travel volumes
and modes.

\leavevmode\vadjust pre{\hypertarget{ref-ine2022}{}}%
Instituto Nacional de Estadísticas, 2022.
\href{https://www.ine.cl/estadisticas}{Estadísticas}.

\leavevmode\vadjust pre{\hypertarget{ref-jain2008gift}{}}%
Jain, J., Lyons, G., 2008. The gift of travel time. Journal of transport
geography 16, 81--89.

\leavevmode\vadjust pre{\hypertarget{ref-kalatian2016travel}{}}%
Kalatian, A., Shafahi, Y., 2016. Travel mode detection exploiting
cellular network data, in: MATEC Web of Conferences. EDP Sciences, p.
03008.

\leavevmode\vadjust pre{\hypertarget{ref-karlaftis2004dea}{}}%
Karlaftis, M.G., 2004. A DEA approach for evaluating the efficiency and
effectiveness of urban transit systems. European Journal of Operational
Research 152, 354--364.

\leavevmode\vadjust pre{\hypertarget{ref-kuhnimhof2006users}{}}%
Kuhnimhof, T., Chlond, B., Von Der Ruhren, S., 2006. Users of transport
modes and multimodal travel behavior: Steps toward understanding
travelers' options and choices. Transportation research record 1985,
40--48.

\leavevmode\vadjust pre{\hypertarget{ref-lee2020millennials}{}}%
Lee, Y., Circella, G., Mokhtarian, P.L., Guhathakurta, S., 2020. Are
millennials more multimodal? A latent-class cluster analysis with
attitudes and preferences among millennial and generation x commuters in
california. Transportation 47, 2505--2528.

\leavevmode\vadjust pre{\hypertarget{ref-li2017estimating}{}}%
Li, G., Chen, C.-J., Peng, W.-C., Yi, C.-W., 2017. Estimating crowd flow
and crowd density from cellular data for mass rapid transit, in:
Proceedings of the 6th International Workshop on Urban Computing,
Halifax, NS, Canada. pp. 18--30.

\leavevmode\vadjust pre{\hypertarget{ref-li2018review}{}}%
Li, Y., Wu, F.-X., Ngom, A., 2018. A review on machine learning
principles for multi-view biological data integration. Briefings in
bioinformatics 19, 325--340.

\leavevmode\vadjust pre{\hypertarget{ref-meta2021physiology}{}}%
Meta, I., Cucchietti, F.M., Navarro-Mateu, D., Graells-Garrido, E.,
Guallart, V., 2022. A physiology-inspired framework for holistic city
simulations. Cities 126, 103553.
\url{https://doi.org/10.1016/j.cities.2021.103553}

\leavevmode\vadjust pre{\hypertarget{ref-casen2020}{}}%
Ministerio de Desarrollo Social y Familia, 2020.
\href{http://observatorio.ministeriodesarrollosocial.gob.cl/encuesta-casen-en-pandemia-2020}{Encuesta
{CASEN} en {P}andemia 2020}.

\leavevmode\vadjust pre{\hypertarget{ref-molin2016multimodal}{}}%
Molin, E., Mokhtarian, P., Kroesen, M., 2016. Multimodal travel groups
and attitudes: A latent class cluster analysis of dutch travelers.
Transportation Research Part A: Policy and Practice 83, 14--29.

\leavevmode\vadjust pre{\hypertarget{ref-monroe2008fightin}{}}%
Monroe, B.L., Colaresi, M.P., Quinn, K.M., 2008. Fightin'words: Lexical
feature selection and evaluation for identifying the content of
political conflict. Political Analysis 16, 372--403.

\leavevmode\vadjust pre{\hypertarget{ref-munizaga2020fare}{}}%
Munizaga, M.A., Gschwender, A., Gallegos, N., 2020. Fare evasion
correction for smartcard-based origin-destination matrices.
Transportation Research Part A: Policy and Practice 141, 307--322.

\leavevmode\vadjust pre{\hypertarget{ref-munizaga2012estimation}{}}%
Munizaga, M.A., Palma, C., 2012. Estimation of a disaggregate multimodal
public transport origin--destination matrix from passive smartcard data
from {S}antiago, {C}hile. Transportation Research Part C: Emerging
Technologies 24, 9--18.

\leavevmode\vadjust pre{\hypertarget{ref-munoz2015encuesta}{}}%
Muñoz, V., Thomas, A., Navarrete, C., Contreras, R., 2015. Encuesta
origen-destino de {S}antiago 2012: Resultados y validaciones. Estudios
de Transporte 19.

\leavevmode\vadjust pre{\hypertarget{ref-pappalardo2023dataset}{}}%
Pappalardo, L., Cornacchia, G., Navarro, V., Bravo, L., Ferres, L.,
2023. A dataset to assess mobility changes in chile following local
quarantines. Scientific Data 10, 6.

\leavevmode\vadjust pre{\hypertarget{ref-phithakkitnukoon2017inferring}{}}%
Phithakkitnukoon, S., Sukhvibul, T., Demissie, M., Smoreda, Z.,
Natwichai, J., Bento, C., 2017. Inferring social influence in transport
mode choice using mobile phone data. EPJ Data Science 6, 1--29.

\leavevmode\vadjust pre{\hypertarget{ref-pineda2019travel}{}}%
Pineda, C., Lira, B.M., 2019. Travel time savings perception and
well-being through public transport projects: The case of {M}etro de
{S}antiago. Urban Science 3, 35.

\leavevmode\vadjust pre{\hypertarget{ref-qu2015transportation}{}}%
Qu, Y., Gong, H., Wang, P., 2015. Transportation mode split with mobile
phone data, in: Intelligent Transportation Systems (ITSC), 2015 IEEE
18th International Conference On. IEEE, pp. 285--289.

\leavevmode\vadjust pre{\hypertarget{ref-rodriguez2014measuring}{}}%
Rodríguez-Núñez, E., García-Palomares, J.C., 2014. Measuring the
vulnerability of public transport networks. Journal of transport
geography 35, 50--63.

\leavevmode\vadjust pre{\hypertarget{ref-rowe2023big}{}}%
Rowe, F., 2023. Big data, in: Concise Encyclopedia of Human Geography.
Edward Elgar Publishing, pp. 42--47.

\leavevmode\vadjust pre{\hypertarget{ref-rowe2022using}{}}%
Rowe, F., 2022. Using digital footprint data to monitor human mobility
and support rapid humanitarian responses. Regional Studies, Regional
Science 9, 665--668.

\leavevmode\vadjust pre{\hypertarget{ref-roweetal2022}{}}%
Rowe, F., Arribas-Bel, D., Calafiore, A., MacDonald, J., Samardzhiev,
K., Fleischmann, M., 2022. Mobility data in urban science. The Alan
Turing Institute.

\leavevmode\vadjust pre{\hypertarget{ref-rowe2020drivers}{}}%
Rowe, F., Bell, M., 2020. The drivers of long-distance commuting in
{C}hile: The role of the spatial distribution of economic activities,
in: Population Change and Impacts in Asia and the Pacific. Springer, pp.
123--149.

\leavevmode\vadjust pre{\hypertarget{ref-schlaich2010generating}{}}%
Schlaich, J., Otterstätter, T., Friedrich, M., others, 2010. Generating
trajectories from mobile phone data, in: Proceedings of the 89th Annual
Meeting Compendium of Papers, Transportation Research Board of the
National Academies.

\leavevmode\vadjust pre{\hypertarget{ref-sectra2012informe}{}}%
SECTRA, 2015. Informe ejecutivo encuesta origen-destino de viajes
{S}antiago.

\leavevmode\vadjust pre{\hypertarget{ref-semanjski2017spatial}{}}%
Semanjski, I., Gautama, S., Ahas, R., Witlox, F., 2017. Spatial context
mining approach for transport mode recognition from mobile sensed big
data. Computers, Environment and Urban Systems 66, 38--52.

\leavevmode\vadjust pre{\hypertarget{ref-smoreda2013spatiotemporal}{}}%
Smoreda, Z., Olteanu-Raimond, A.-M., Couronné, T., 2013. Spatiotemporal
data from mobile phones for personal mobility assessment, in: Transport
Survey Methods: Best Practice for Decision Making. Emerald Group
Publishing Limited.

\leavevmode\vadjust pre{\hypertarget{ref-somma2021no}{}}%
Somma, N.M., Bargsted, M., Disi Pavlic, R., Medel, R.M., 2021. No water
in the oasis: The {C}hilean {S}pring of 2019--2020. Social Movement
Studies 20, 495--502.

\leavevmode\vadjust pre{\hypertarget{ref-suazo2019displacement}{}}%
Suazo-Vecino, G., Muñoz, J.C., Fuentes Arce, L., 2019. The displacement
of {S}antiago de {C}hile's downtown during 1990--2015: {T}ravel time
effects on eradicated population. Sustainability 12, 289.

\leavevmode\vadjust pre{\hypertarget{ref-pyrosm}{}}%
Tenkanen, H., contributors, pyrosm, 2022. {pyrosm}: {O}pen{S}treet{M}ap
{PBF} data parser for {P}ython.

\leavevmode\vadjust pre{\hypertarget{ref-thomson1998reflections}{}}%
Thomson, J.M., 1998. Reflections on the economics of traffic congestion.
Journal of transport economics and policy 93--112.

\leavevmode\vadjust pre{\hypertarget{ref-tirachini2017estimation}{}}%
Tirachini, A., Hurtubia, R., Dekker, T., Daziano, R.A., 2017. Estimation
of crowding discomfort in public transport: Results from {S}antiago de
{C}hile. Transportation Research Part A: Policy and Practice 103,
311--326.

\leavevmode\vadjust pre{\hypertarget{ref-tirachini2019ride}{}}%
Tirachini, A., Río, M. del, 2019. Ride-hailing in {S}antiago de {C}hile:
Users' characterisation and effects on travel behaviour. Transport
Policy 82, 46--57.

\leavevmode\vadjust pre{\hypertarget{ref-urquieta2019metro}{}}%
Urquieta Ch., C., 2019.
\href{https://www.ciperchile.cl/2019/10/24/dura-perdida-para-el-metro-no-tiene-seguros-para-estaciones-ni-trenes/}{Dura
pérdida para el metro: No tiene seguros para estaciones ni trenes}.
CIPER.

\leavevmode\vadjust pre{\hypertarget{ref-vuchic2007urban}{}}%
Vuchic, V.R., 2007. Urban transit systems and technology. John Wiley \&
Sons.

\leavevmode\vadjust pre{\hypertarget{ref-wang2010transportation}{}}%
Wang, H., Calabrese, F., Di Lorenzo, G., Ratti, C., 2010. Transportation
mode inference from anonymized and aggregated mobile phone call detail
records, in: Intelligent Transportation Systems (ITSC), 2010 13th
International IEEE Conference On. IEEE, pp. 318--323.

\leavevmode\vadjust pre{\hypertarget{ref-wu2013studying}{}}%
Wu, W., Cheu, E.Y., Feng, Y., Le, D.N., Yap, G.E., Li, X., 2013.
Studying intercity travels and traffic using cellular network data.
Mobile Phone Data for Development: Net Mob 2013.

\leavevmode\vadjust pre{\hypertarget{ref-zannat2019emerging}{}}%
Zannat, K.E., Choudhury, C.F., 2019. Emerging big data sources for
public transport planning: A systematic review on current state of art
and future research directions. Journal of the Indian Institute of
Science 99, 601--619.

\leavevmode\vadjust pre{\hypertarget{ref-zhang2019using}{}}%
Zhang, L., Pfoser, D., 2019. Using OpenStreetMap point-of-interest data
to model urban change---a feasibility study. PloS one 14, e0212606.

\leavevmode\vadjust pre{\hypertarget{ref-vzitnik2014data}{}}%
Žitnik, M., Zupan, B., 2014. Data fusion by matrix factorization. IEEE
transactions on pattern analysis and machine intelligence 37, 41--53.

\end{CSLReferences}

% DON'T EDIT. If "endfloat" option is enabled all floats appear before appendices
\if@endfloat\clearpage\processdelayedfloats\clearpage\fi

%%%%%%%%%%%%%%%%%%%%%%%%%%%%%%%%%%%%%%%%%%%%%%%%%%%%%%%%%%%%
%%% SUPPLEMENTARY MATERIAL / APPENDICES
%%%%%%%%%%%%%%%%%%%%%%%%%%%%%%%%%%%%%%%%%%%%%%%%%%%%%%%%%%%%
%% Sadly, we can't use floats in the appendix boxes. So they don't "float", but use \captionof{figure}{...} and \captionof{table}{...} to get them properly caption.
%\begin{appendix}

%\begin{appendixbox}\label{app:ttt}
%    \input{src/supplementary/appendices.tex}
%\end{appendixbox}

%\begin{appendixbox}
%    \input{src/supplementary/resources.tex}
%\end{appendixbox}

%\end{appendix}

%%%%%%%%%%%%%%%%%%%%%%%%%%%%%%%%%%%%%%%%%%%%%%%%%%%%%%%%%%%%
%%% ARTICLE END
%%%%%%%%%%%%%%%%%%%%%%%%%%%%%%%%%%%%%%%%%%%%%%%%%%%%%%%%%%%%

\end{document}